\documentclass[runningheads]{llncs}
\usepackage[T1]{fontenc}
\usepackage{graphicx}
\usepackage{stmaryrd}
\usepackage{mathbbol}
\usepackage{times}
\usepackage{setspace}

\usepackage{latexsym,amsmath,epsfig,amssymb,graphics,tabularx}
\usepackage{color}
\usepackage{prftree}
\usepackage{tikz}
\usepackage{algorithm}      
\usepackage{algpseudocode}  
\algrenewcommand\algorithmicrequire{\textbf{Input:}}
\algrenewcommand\algorithmicensure{\textbf{Output:}}

\usepackage{mathbbol}
\newcommand{\pskip}{\textmd{skip}}

\newcommand{\pwait}{\textrm{wait}}
\newcommand{\exempt}[4]{#1 \unrhd \talloblong_{#2} (#3 \rightarrow #4)}

\newcommand{\IFE}[3]{\textrm{if}\ #1\ \textrm{then}\ #2\ \textrm{else}\ #3}

\newcommand{\evo}[3]{\langle \overrightarrow{\dot{#1}}=\overrightarrow{#2}\& #3\rangle}
\newcommand{\rdy}{\mathit{rdy}}
\newcommand{\chop}{\smallfrown}
\newcommand{\compat}{\operatorname{compat}}

\newcommand{\cm}{\mathit{cm}}
\newcommand{\dL}{d\mathcal{L}}
\newcommand{\oomit}[1]{}

\begin{document}
\title{HHLPar: Automated Theorem Prover for Parallel Hybrid Communicating Sequential Processes}
\titlerunning{HHLPar}
%
\author{Xiangyu Jin\inst{1,2}\and
Bohua Zhan\inst{3} \and
Shuling Wang\inst{4,2}\thanks{Corresponding author.}\and
Naijun Zhan\inst{5}
}
\institute{}
\institute{Key Laboratory of System Software and State Key Lab. of Computer Science, Institute of Software Chinese Academy of Sciences, Beijing, China \\
\email{jinxy@ios.ac.cn}
\and
University of Chinese Academy of Sciences \and
Huawei Technologies Co., Ltd., Beijing, China\\
\email{zhanbohua@huawei.com} \and
National Key Laboratory of Space Integrated Information System, Institute of Software Chinese Academy of Sciences, Beijing, China\\
\email{wangsl@ios.ac.cn}
\and 
School of Computer Science, Peking University, Beijing, China\\
\email{njzhan@pku.edu.cn}
}
\maketitle              
\begin{abstract}
We introduce HHLPar, a tool for verifying hybrid systems modeled in Hybrid Communicating Sequential Processes (HCSP). HHLPar is based on a Hybrid Hoare Logic for HCSP, which enables reasoning about both the continuous-time properties of differential equations and the communication and parallel composition of HCSP processes. This is achieved through the use of specialized trace assertions and their synchronization. The logic has been formalized and proven sound in Isabelle/HOL, providing a reliable foundation for the verification.
HHLPar implements the logic in Python and supports automated verification: On one hand, it provides functions for symbolically decomposing HCSP processes, generating assertions for  individual sequential processes, and then composing them via synchronization to obtain the final specification for the entire parallel HCSP process; On the other hand, it is integrated with external solvers for handling differential equations and real arithmetic properties. The resulting assertions are sufficiently expressive to deduce both the state properties at termination and the continuous-time invariants maintained throughout the execution of processes, which are critical for ensuring system safety. Finally, we present the main issues related to the implementation of HHLPar and demonstrate its applicability through a case study involving a simplified cruise control system.

\end{abstract}
\section{Introduction}

Hybrid systems involve complex interactions between continuous-time evolving physical processes and discrete control systems. In networked applications such as cyber-physical systems, communication and parallel composition play a critical role in enabling interactions among distributed components, to facilitate the coordination of concurrent behaviors and the exchange of data across subsystems. However, ensuring the safety of such systems is highly challenging due to their inherent complexity, which stems from the interplay of continuous dynamics, discrete transitions, and the need for synchronization  between parallel components. Formal verification has been widely recognized in both academic community and industry as an important approach to ensure correctness of hybrid  systems. Especially, a verification tool that is sound and capable of producing trustworthy results and meanwhile supporting automation in verification process  is essential for the practical design of safety-critical systems.   

There are two mainstream verification techniques of hybrid systems: model checking and deductive verification. Model checking verifies a system model, typically represented as hybrid automata~\cite{Alur1993}, by exhaustively computing and checking all reachable system states. However, this approach faces intrinsic challenges due to the infinite state domains and the increasing complexity of hybrid systems.  On the other hand, deductive verification  conducts proof via logical reasoning by induction on system models and reasons about continuous evolution represented as ordinary differential equations (ODEs) with the help of differential invariants~\cite{PlatzerClarke08,LLQZ10,LiuEmsoft}. A prerequisite for deductive verification of hybrid systems is to have a compositional modelling language for hybrid    systems and meanwhile a specification logic for reasoning about the formal models such that the verification of a complex system can be reduced to the verification of decomposed components of the system. Differential dynamic logic ($\dL$)~\cite{DBLP:journals/jar/Platzer08,Platzer,bohrer2017,DBLP:books/sp/Platzer18} is a first-order dynamic logic proposed for specifying and verifying hybrid systems modelled as hybrid programs. Its soundness has been proved in Isabelle/HOL and Coq in ~\cite{DBLP:conf/cpp/BohrerRVVP17}.  Its prover KeYmaera~\cite{DBLP:conf/cade/PlatzerQ08} supports automatic proof search of rules of $\dL$ and integrates with computer algebra tools for solving differential equations and real arithmetic  formulas. Its successor KeYmaera X~\cite{DBLP:conf/cade/FultonMQVP15} enhances automation and provides stronger soundness guarantees through a small, trusted prover kernel. However, $\dL$ lacks direct support for communication and parallel composition, which are ubiquitous in practical cyber-physical systems. The verification of hybrid systems with communication and parallel composition poses additional challenges due to the need to account for concurrent interactions, synchronization  and the resulting complex, non-deterministic behaviors arising from distributed components.

\begin{figure}
    \centering    \includegraphics[scale=0.23]{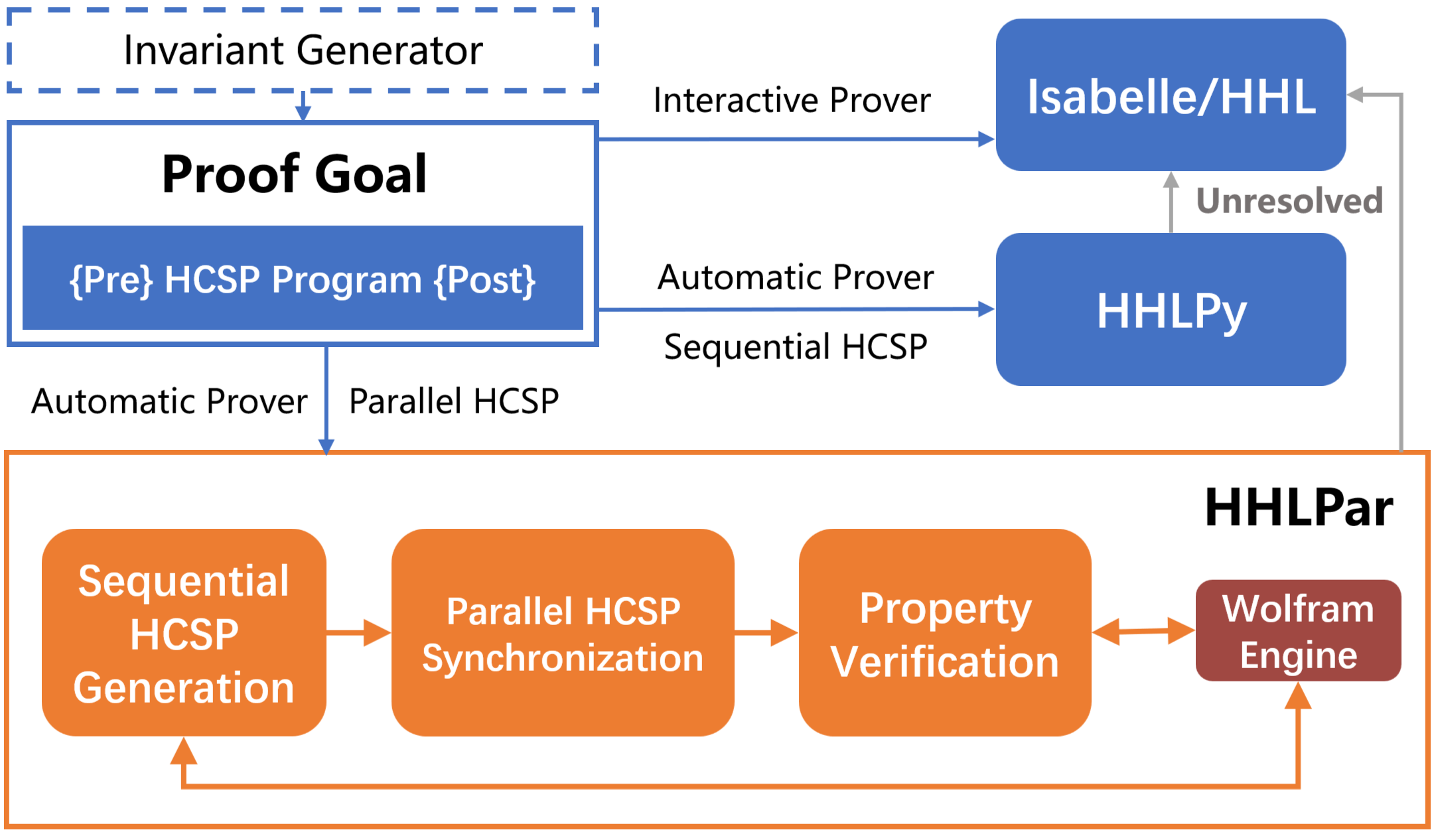}    
    \vspace{-5pt}
    \caption{Architecture of HHLProver.}
    \label{HHLProver}
\end{figure}

Hybrid CSP (HCSP)~\cite{Jifeng:1994,Zhou:1996} extends Hoare's CSP~\cite{Hoare85}  by including  ODEs to model continuous dynamics. It leverages the communication and parallel composition features of CSP to enable the flexible interactions between continuous physical processes and discrete control systems. The specification logic and verification of HCSP have been studied by extending the classical  Hoare logic to handle both continuous evolution and communication based parallel composition. One line of the work~\cite{LLQZ10,WZG12} utilizes Duration Calculus (DC), which is an interval-based temporal logic with binary modality chop and was extended to specify continuous-time properties, but the DC-based reasoning system is quite complicated and in consequence the tool support for verifying HCSP under
this approach is limited to interactive theorem proving in Isabelle/HOL~\cite{WZZ15}, which imposes a significant proof burden on users. To overcome these limitations,  an alternative Hybrid Hoare Logic (HHL) was developed by introducing trace-based assertions into first-order logic~\cite{DBLP:journals/corr/abs-2303-15020}.  This logic proposes traces composed of  both communication and continuous-time events, and handles parallel composition of processes through  trace synchronization. 
Building on this logic,   the HHL prover was implemented, as illustrated in Fig.~\ref{HHLProver}, providing a more automated and user-friendly verification tool for HCSP.

As shown by Fig.~\ref{HHLProver}, the HHL prover comprises four parts: an Invariant Generator for synthesizing differential invariants of ODEs  and supplying them to other modules;  HHLPy~\cite{DBLP:conf/fm/ShengBZ23}, an automatic verifier for verifying sequential HCSP, particularly ODEs, based on differential invariants; HHLPar, an automatic verifier for HCSP with communication and concurrency; and Isabelle/HHL, an interactive theorem prover  for HHL. Both HHLPy and HHLPar are designed to automate verification, while unproven conditions  are passed to  the interactive mode of HHL prover, i.e. Isabelle/HHL. HCSP captures flexible interactions among multiple processes via communication, which is typical in cyber-physical systems, while HHLPy is restricted to single, sequential processes — this limitation motivates our work.

In this paper, we present  HHLPar, the automated theorem prover for HCSP in concurrent setting, including its assertions, inference rules and implementation.
HHLPar builds upon the HHL in~\cite{DBLP:journals/corr/abs-2303-15020} but differs in several key aspects.  The HHL in~\cite{DBLP:journals/corr/abs-2303-15020} defines a generalized trace-based logic and a weakest precondition-style proof system, which is proved to be relative complete and  very expressive, but faces difficulty in automating the verification of parallel composition.  Instead, HHLPar  proposes  an explicit assertion language for specifying traces,  and provides a set of inference rules for constructing assertions of sequential  processes and a set of synchronization rules for constructing assertions of parallel processes constituting their specification. This constructive style logic 
sacrifices the relative completeness of the original logic but enables the automation of  HCSP reasoning.
The soundness of these inference rules underlying HHLPar has been formally verified in Isabelle/HOL. 
Meanwhile, we implemented the algorithms for operating assertions in Python, which supports the automated theorem proving of HCSP processes by performing symbolic decomposition and reasoning based on the logic's inference rules. It also inherits HHLPy's integration with the external Wolfram Engine for automatically solving ODEs and reasoning about logical formulas. 

The assertions in specification generated by HHLPar are sufficiently expressive to describe the behavior of processes. Also, it is strong enough to enable the derivation of logical formula properties over process variables. In this paper we have developed a set of inference rules specifically for deriving two different forms of properties from the generated assertions automatically. The first class is properties of final states at termination which is also a concern of classical Hoare logic and HHLPy. The second class is continuous-time invariants held throughout the execution which ensure that the system meets the requirements over all continuous time intervals. These properties are crucial for assessing system safety. To demonstrate the usability of HHLPar, we applied it to verify a simplified cruise control system, successfully automating the verification of its safety requirement.

After reviewing the related work, the remainder of the paper is structured as follows. Sect.~\ref{sec:hcsp} provides a brief overview of HCSP. 
Sect.~\ref{Assertions} introduce the assertions we proposed and its corresponding specification modified from HHL. 
Sect.~\ref{SequentialHCSP} and Sect.~\ref{ParallelHCSP} introduce inference rules of how to construct assertions in specification for both sequential and parallel HCSP, respectively. 
Sect.~\ref{Propertyv} gives the rules for proving properties of specific forms from assertions.
Sect.~\ref{sec:case} discusses the key implementation aspects in both Isabelle/HOL and HHLPar, and demonstrates the application of HHLPar through a case study. The accompanying code, including the formalization and soundness proof of the logic in Isabelle/HOL, the Python implementation and the case study, is available at https://github.com/AgHHL/gHHL2024.git.

\subsection{Related Work}

Model checking tools of hybrid systems endeavor to compute reachable states of continuous dynamics  efficiently in an algorithmic approach, by achieving high scalability while maintaining high accuracy, e.g. the representative PHAVer~\cite{DBLP:journals/sttt/Frehse08} for linear hybrid automata, HSolver~\cite{DBLP:journals/tecs/RatschanS07} and SpaceEx~\cite{DBLP:conf/cav/FrehseGDCRLRGDM11} for both linear and non-linear dynamics. Deduction verification tools are developed upon program logics and conduct proofs via theorem proving.  KeYmaera~\cite{DBLP:conf/cade/PlatzerQ08} and its successor KeYmaera X~\cite{DBLP:conf/cade/FultonMQVP15} are automated and interactive theorem provers built upon differential dynamic logic ($\dL$)~\cite{DBLP:journals/jar/Platzer08,Platzer,DBLP:books/sp/Platzer18},  which proposes a complete set of rules~\cite{DBLP:journals/jar/Platzer17,DBLP:journals/jacm/PlatzerT20} for reasoning about continuous dynamics such as differential invariants, differential weakening, differential cut, and differential ghosts. Both the tools combine deductive reasoning of $\dL$, real algebraic and computer algebraic provers for automated verification. 
Foster et al.~\cite{DBLP:conf/fm/FosterMGS21} proposed a semantic verification framework for hybrid systems using the Isabelle/HOL proof assistant and then extended it to IsaVODEs~\cite{DBLP:journals/jar/MuniveFGSLH24}.
The related work on specification and verification of HCSP have been discussed in the introduction. In contrast, HHLPar extends HHLPy~\cite{DBLP:conf/fm/ShengBZ23} to support the parallel fragment of HCSP, encompassing communication, parallel composition and continuous evolution. HHLPar inherits HHLPy's integration with external solvers for real arithmetic and ODEs, and further enables automated deductive verification of communication and parallel composition through specialized assertions and synchronization. Both HHLPy and HHLPar are integrated to HHL prover in order to improve its automation, as indicated in Fig.~\ref{HHLProver}.

\section{An Overview of HCSP}
\label{sec:hcsp}

As an extension of Communicating Sequential Processes (CSP~\cite{Hoare85}), Hybrid CSP (HCSP) is a formal modeling language for hybrid systems. It introduces Ordinary Differential Equations (ODEs) to model continuous evolution and interrupts. In HCSP, communication is the sole mechanism for data exchange between processes, and shared variables among parallel processes are explicitly prohibited.
This section is extracted  from \cite{DBLP:journals/corr/abs-2303-15020}, which plays  the foundation of the logic in this paper. For self-containedness,  we provide a brief overview.
\paragraph{Syntax.}
Below, we present the syntax for HCSP. Here $c$ and $c_i$ denote sequential processes, while $pc$ and 
 $pc_i$ denote parallel processes. $\dot{x}$ represents the first-order derivative of $x$ w.r.t. time, $\overrightarrow{x}$ (resp. $\overrightarrow{e}$) denotes a vector of variables (expressions). $ch$ refers to a channel name, and $ch_i*$ denotes  either an input event $ch_i?x$ or output event $ch_i!e$. $L$ is a non-empty set of indices, $cs$ is a set of channel names. $B$ and $e$ represent Boolean and arithmetic expressions, respectively.
 \vspace{-5pt}
{\small
\[
\begin{array}{lll}  
	c  & ::= & 
	\pskip \mid
	x := e \mid 
	ch?x \mid 
	ch!e \mid 
	c_1 \sqcup c_2 \mid
	c_1; c_2 \mid c^* \mid
	\IFE{B}{c_1}{c_2} \mid \\
	& &
	\evo{x}{e}{B} \mid \pwait\ e \mid\exempt{\evo{x}{e}{B\propto c}}{i\in L}{ch_i*}{c_i} \\
	pc &::=& c \mid pc_1\|_{cs} pc_2 
\end{array} 
\vspace{-5pt}
\]
}
\par\noindent
The input $ch?x$ receives a value through channel $ch$ and assigns it to variable $x$, while the output $ch!e$ sends the value of $e$ through $ch$. Both statements may block, waiting for the corresponding dual party to be ready.
The continuous evolution $\evo{x}{e}{B}$ evolves continuously according to the given ODE $\overrightarrow{\dot{x}}=\overrightarrow{e}$ as long
as the open \emph{domain} $B$ holds, and terminates whenever $B$ becomes false. 
The wait statement $\pwait\ e$ keeps variables unchanged except that a period of time determined by $e$ progresses. 
Communication interruption 	$\exempt{\evo{x}{e}{B\propto c}}{i\in L}{ch_i*}{c_i}$ evolves according to the ODE $\overrightarrow{\dot{x}}=\overrightarrow{e}$ until it is preempted by one of the communication events $ch_i*$, followed by the corresponding $c_i$; or until it violated the domain condition $B$, followed by the execution of $c$.  
The parallel composition $pc_1\|_{cs} pc_2$ executes $pc_1$ and $pc_2$  independently, except that all communication events over the common channels in $cs$ are synchronized  between $pc_1$ and $pc_2$. No same channel direction (e.g. $ch!$) occurs in both $pc_1$ and $pc_2$.
The meaning of other statements such as assignment, internal choice, sequential
composition, and so on, follow their standard definitions. 

The following example models a moving vehicle operating in parallel with its discrete controller.  The vehicle's motion is governed by an ODE, where $s$ represents the trajectory, $v$ the velocity and $a$ the acceleration. Every $d$ time units, the continuous evolution is interrupted by the controller. During each interruption, the controller senses the trajectory and the velocity of the vehicle through input $p2c?x$, computes the new acceleration and sends  it to the vehicle via $c2p!\textit{contl(x)}$. The vehicle then follows this updated acceleration in the next time period.  
\vspace{-5pt}
{ \small
 \[
\begin{array}{ll}
  (\exempt{\dot{s} = v, \dot{v} = a}{}{p2c!(s, v)}{c2p?a)}^*  
\|  (\pwait\ d; p2c?x; c2p!contl(x))^*
\end{array}
\vspace{-5pt}
\] }
\paragraph{Semantics}
Fig.~\ref{fig:big-step} presents part of the big-step semantics of HCSP,  defined as a set of transition rules. Each transition takes the form $(c,s)\Rightarrow (s',\textit{tr})$,  
indicating that   $c$ carries initial state $s$ to final state $s'$, producing a trace $\textit{tr}$. Here states $s, s' \in \textit{Vars} \rightarrow\textit{Values}$ are mappings from variables to values.  A trace $tr$ is an ordered sequence of events  generated during the execution of an HCSP process. It can be an empty trace $\epsilon$, a single event, or the concatenation $tr_1^\chop tr_2$ of two traces $tr_1$ and $tr_2$, defined recursively. An event describes an observable step in the behavior of a process. There are two types of events:   A \emph{communication event} $\langle ch\triangleright , v\rangle$, where $\triangleright$ is $?$ or $!$, indicating input and  output, and $v$ is a  value transmitted during the communication; a \emph{continuous event}  $\langle d, \overrightarrow{p}, \rdy\rangle$, where $d$ is a positive value specifying the duration of this event, $\overrightarrow{p}$ a continuous function from $[0,d]$ to states, describing the evolution of states over time, 
 and $\rdy$ is the set of channels that are waiting for communication during this duration. 
\begin{figure*} \centering  
{\small 
 \begin{eqnarray*} 
&\prftree[r]{Out-1 }{(ch!e, s) \Rightarrow (s, \langle ch!, s(e)\rangle)}
		\quad
		\prftree[r]{Out-2}{(ch!e, s) \Rightarrow (s, \langle d, I_s, \{ch!\}\rangle ^\chop \langle ch!, s(e)\rangle)} 
    &\\
   	& 	\prftree[r]{Cont}{
		\begin{array}{cc}
			 \forall t\in[0,d).\,s[\overrightarrow{x}\mapsto \overrightarrow{p}(t)](B) \quad \neg s[\overrightarrow{x}\mapsto \overrightarrow{p}(d)](B)
	\end{array}}
	{(\evo{x}{e}{B}, s) \Rightarrow (s[\overrightarrow{x}\mapsto \overrightarrow{p}(d)], \langle d, \overrightarrow{p}, \{\}\rangle) }
 & \\
 &\prftree[r]{Int-1}{
		\begin{array}{cc}
\forall t\in[0,d).\,s[\overrightarrow{x}\mapsto \overrightarrow{p}(t)](B) \quad  
   i\in L\quad ch_i*=ch!e\quad (c_i,s[\overrightarrow{x}\mapsto \overrightarrow{p}(d)])\Rightarrow (s',\textit{tr})
	\end{array}}
	{\begin{array}{cc}
	     (\exempt{\evo{x}{e}{B\propto c}}{i\in L}{ch_i*}{c_i}, s) \Rightarrow  \\
	     (s',
		\langle d, \overrightarrow{p}, \rdy(\cup_{i\in L} ch_i*)\rangle^\chop \langle ch!,s[\overrightarrow{x}\mapsto \overrightarrow{p}(d)](e)\rangle ^\chop \textit{tr}) 
	\end{array}
  } &\\
  &\prftree[r]{Int-2}{
		\begin{array}{cc}
			 \forall t\in[0,d).\,s[\overrightarrow{x}\mapsto \overrightarrow{p}(t)](B) \quad \neg s[\overrightarrow{x}\mapsto \overrightarrow{p}(d)](B) \quad 
    (c,s[\overrightarrow{x}\mapsto \overrightarrow{p}(d)])\Rightarrow(s',tr)
	\end{array}}
	{\begin{array}{cc}
	     (\exempt{\evo{x}{e}{B\propto c}}{i\in L}{ch_i*}{c_i}, s) \Rightarrow  \\
	     (s',\langle d,\overrightarrow{p},\rdy(\cup_{i\in L} ch_i*)\rangle^\chop tr) 
	\end{array}}
 &\\
 &\prftree[r]{Par}{(c_1, s_1) \Rightarrow (s_1', tr_1)}{(c_2, s_2) \Rightarrow (s_2', tr_2)}{tr_1\|_{cs}tr_2 \Downarrow tr}
{(c_1 \|_{cs} c_2, s_1\uplus s_2) \Rightarrow (s_1'\uplus s_2', tr)}&
 \end{eqnarray*}
}  
\vspace{-25pt}
\caption{Part of big-step  semantics of HCSP}
\vspace{-20pt}
\label{fig:big-step}  
\end{figure*}  

Rules (Out-1) and (Out-2) define two cases for communication: one where the communication occurs immediately, and another where it occurs after a delay of 
$d$ time units. During the waiting period, $I_s$
represents an identity function that maps time to the initial state. Rule (Cont) defines the behavior of the continuous evolution, which terminates after time $d$ due to the violation of domain $B$. This results in a continuous event with duration $d$ and function $\overrightarrow{p}$, where 
$\overrightarrow{p}$ is a solution of the ODE $\overrightarrow{\dot{x}}=\overrightarrow{e}$ satisfying the initial condition $\overrightarrow{p}(0)=s(\overrightarrow{x})$. Rule (int-1) defines that  the ODE is interrupted after $d>0$ time duration, by the occurrence of a communication over channel $ch$, and then the subsequent process $c_i$ is executed; Rule (int-2) defines that the ODE terminates due to the violation of $B$, without any communication among $\{ch_i\}$ being able to occur, and then the subsequent process $c$ is executed. Other similar cases, e.g. interruption by an input event,  are not listed here. Rule (Par) defines the semantics of the parallel composition, which results in the disjoint union of the states (denoted by $s_1\uplus s_2$) and the synchronization of the traces (denoted by $tr_1\|_{cs}tr_2 \Downarrow tr$), of the two respective processes. 

Especially, the trace synchronization relation $tr_1 \|_{cs} tr_2 \Downarrow \textit{tr}$ can be derived according to the structures of traces $tr_1$ and $tr_2$. 
Part of the derivation rules is given below. 
 An output event synchronizes with the corresponding input event (SyncIO). When an external communication event occurs on one side, it does not need to synchronize with the other side (NoSyncIO); When both sides are continuous events, then the continuous events of the same length  will synchronize if they have compatible ready sets (SWait), denoted by $\compat$, meaning that no input and output along a same channel occur simultaneously in the two ready sets (otherwise the corresponding communication must occur immediately). 
\vspace{-5pt}
{\small
\[
\begin{array}{c}
\prftree[r]{SyncIO}{ch\in cs}{tr_1\|_{cs} tr_2 \Downarrow \textit{tr}}
{\langle ch!,v\rangle^\chop tr_1 \|_{cs} \langle ch?,v\rangle ^\chop tr_2 \Downarrow \textit{tr}} \quad 
\prftree[r]{NoSyncIO}{ch\notin cs}{tr_1\|_{cs} tr_2 \Downarrow \textit{tr}}
{\langle ch\triangleright,v\rangle^\chop tr_1 \|_{cs} tr_2 \Downarrow \langle ch\triangleright,v\rangle^\chop \textit{tr}}
\\
\prftree[r]{SWait}
{tr_1\|_{cs} tr_2 \Downarrow \textit{tr}}
{\compat(\rdy_1,\rdy_2) \quad d >0 }
{
\begin{array}{c}
    \langle d,\overrightarrow{p}_1,\rdy_1\rangle^\chop tr_1 \|_{cs} \langle d,\overrightarrow{p}_2,\rdy_2\rangle^\chop tr_2 \Downarrow  
    \langle d, \overrightarrow{p}_1\uplus \overrightarrow{p}_2,(\rdy_1\cup \rdy_2)-cs\rangle^\chop \textit{tr} 
\end{array}
}
\end{array}
\]}
\vspace{-15pt}

\oomit{
\paragraph{\textbf{Hoare Triples}}
In trace-based hybrid Hoare logic, the specification of a given process $c$ takes the form of Hoare triple $\{P\}\,c\,\{Q\}$, where $P$ and $Q$ are predicates on state and trace. The validity of this triple is defined in terms of big-step  semantics as follows:
{\small
\[
\models \{P\}\,c\,\{Q\} \triangleq
\forall s_1\ s_2\ tr\ tr'.\, (s_1,tr)\models P 
\longrightarrow (c,s_1)\Rightarrow (s_2,tr')
\longrightarrow (s_2,tr^\chop tr')\models Q
\]}}

\section{Assertions and Specifications}\label{Assertions}
We will introduce an assertion language for explicitly specifying traces,  which serves as the foundation for the inference rules of constructing specifications of HCSP in the following sections. The assertion language, with its explicit syntactic forms, enables automated processing of inference rules for verifying HCSP processes. Building on these assertions, we further propose a novel specification form tailored for HCSP.

\subsection{Syntax and Semantics}
The syntax of the assertion language is defined below: $P, Q$ represent assertions, $cm$ is a list of tuples recording the assertion information for channels, $I$ is a path condition. 
\vspace{-5pt}
{\small
\[
\begin{array}{lll}  
	P,Q  & ::= & 
	\mathsf{true} \mid \mathsf{false} \mid P \bar{\wedge} Q \mid P \bar{\vee} Q \mid \uparrow b \mid P[\overrightarrow{x}:=\overrightarrow{e}] \mid \mathsf{init}\\
 && \mid\mathsf{wait\_in}(I,ch,\{\mathsf{d,v}\Rightarrow P\}) \mid \mathsf{wait\_outv}(I,ch,e,\{\mathsf{d}\Rightarrow P\}) \mid \mathsf{wait}(I,e,\{\mathsf{d}\Rightarrow P\}) \\
 &&\mid \mathsf{interrupt}(I,e,\{\mathsf{d}\Rightarrow P\},\cm) \mid \mathsf{interrupt_\infty}(I,\cm) \mid \mathsf{Rec}\ R.\ P\bar{\vee} F(R)\\
 cm  & ::= & \epsilon \mid  (ch?, \{\mathsf{d,v}\Rightarrow P\}) \cdot cm \mid ( ch!,h, \{\mathsf{d}\Rightarrow P\}) \cdot cm   \\
 I & ::= & \mathsf{id} \mid \overrightarrow{x} \rightarrowtail f(\overrightarrow{x},t) \mid \mathsf{inv} \mid I[\overrightarrow{x}:=\overrightarrow{e}]\mid I_1\uplus I_2
\end{array} 
\vspace{-5pt}
\]}
\par\noindent
where $b$ and $\mathsf{inv}$ are boolean expressions, $e$ is a real expression,   $\{\mathsf{d,v}\Rightarrow P\}$ represents a function mapping from real valued variables $\mathsf{d}$ and $\mathsf{v}$ to assertions($\{\mathsf{d}\Rightarrow P\}$ is similar), for example, $\{\mathsf{d,v}\Rightarrow\mathsf{init}[x:=x+\mathsf{d}][y:=\mathsf{v}]\}$. Here, $\mathsf{d}$ and $\mathsf{v}$ are two special bounded variables introduced to synchronize communication between parallel processes: $\mathsf{d}$ denotes the transmitted value and  $\mathsf{v}$ denotes the time of occurrence, which will be resolved when the dual events in parallel processes synchronize.
$cm$ is a list of tuples or triples recording the communication branches used in interrupt.  $\mathsf{Rec}$  defines a recursive assertion where 
$P$ acts as the guard ensuring the recursion terminates.
Here $F$ is a generator function defined inductively according to the syntax of the assertion language which can be atomic or non-atomic assertion containing a hole indicating the position where a recursion happens. For example, $F(R)$ can be $R[x:=0]$ or $\mathsf{wait}(I,e,\{\mathsf{d}\Rightarrow R[x:=x+1]\})$.

We  first define the semantics of path conditions. 
A path condition $I$ is a predicate interpreted over the starting state, the time and the state, denoted by $(s_0,t,s)\models I$. It describes the relationship between the starting state $s_0$ and the state $s$ at time $t$ during the evolution. As defined by the semantics, $\mathsf{id}$ states that $s$ keeps the same as the initial state $s_0$; $\overrightarrow{x} \rightarrowtail f(\overrightarrow{x},t)$ substitutes $\overrightarrow{x}$ to the value defined by $f$ at time $t$;  $\mathsf{inv}$ means that state $s$ at $t$  satisfies the invariant $\mathsf{inv}$; the substitution  $I[\overrightarrow{x}:=\overrightarrow{e}]$ updates the value of $\overrightarrow{x}$  at initial state to be the one of $\overrightarrow{e}$. Intuitively, we use $\mathsf{id}$ to describe the constant duration and use $f$ and $\mathsf{inv}$ to handle the ODE with explicit solutions or with differential invariants.
\vspace{-5pt} 
{\small\[
    \begin{array}{rcl}
         (s_0,t,s)\models \mathsf{id} &\triangleq& s=s_0  \\
         (s_0,t,s)\models \overrightarrow{x} \rightarrowtail f(\overrightarrow{x},t)&\triangleq& s=s_0[\overrightarrow{x}\mapsto f(s_0(\overrightarrow{x}),t)]\\(s_0,t,s)\models\mathsf{inv}&\triangleq& \mathsf{inv}(s)\\        (s_0,t,s)\models I[\overrightarrow{x}:=\overrightarrow{e}]&\triangleq& 
         (s_0[\overrightarrow{x}\mapsto s_0(\overrightarrow{e})],t,s)\models I\\
         (s_0,t,s)\models I_1\uplus I_2 &\triangleq& \exists\,s_{01}\,s_{02}\,s_{1}\,s_{2}.\, s_0=s_{01}\uplus s_{02}\wedge s=s_{1}\uplus s_{2}\wedge \\&&(s_{01},t,s_1)\models I_1\wedge(s_{02},t,s_2)\models I_2
         
    \end{array}
 \]}
\vspace{-10pt} 


Next, we introduce the semantics of the assertions. An assertion $P$ is interpreted over an initial state, current state and a trace, denoted by $(s_0,s,tr)\models P$. The assertions $\mathsf{true}, \mathsf{false}, P \bar{\wedge} Q, P \bar{\vee} Q $ are defined similar to the propositional logic. $\uparrow b$ lifts a boolean expression on starting state as a boolean assertion, i.e. $b$ holds at the starting state. 
$P[\overrightarrow{x}:=\overrightarrow{e}]$ means that $P$ holds under the starting state updated by assigning $\overrightarrow{x}$ to $\overrightarrow{e}$.
$\mathsf{init}$ means that the state equals starting state and the trace is empty.
\vspace{-5pt} 
{\small
\[
\begin{array}{rcl}


(s_0,s,tr)\models {\uparrow b} &\triangleq& b(s0) \\

(s_0,s,tr)\models P \bar{\wedge} Q &\triangleq& (s_0,s,tr)\models P \wedge (s_0,s,tr)\models Q \\

(s_0,s,tr)\models P \bar{\vee} Q &\triangleq& (s_0,s,tr)\models P \vee (s_0,s,tr)\models Q \\

(s_0,s,tr)\models P[\overrightarrow{x}:=\overrightarrow{e}] &\triangleq& (s_0[\overrightarrow{x}\mapsto s_0(\overrightarrow{e})],s,tr)\models P 
 \\
 (s_0,s,tr)\models \mathsf{init} &\triangleq&  s_0=s \wedge tr=\epsilon 


\end{array}
\]
}
\vspace{-10pt} 
\oomit{
Next, we'll introduce a series of assertions describing modifications to traces.
\begin{definition}
   \[
(s_0,s,tr)\models \mathsf{init} \triangleq s_0=s \wedge tr=\epsilon 
\] 
\end{definition}
The simplest assertion $\mathsf{init}$ means the state equals starting state and the trace is empty.}

We then introduce the semantics of assertions specifying the behavior of  input, output, continuous evolution and interrupt respectively: 
\vspace{-5pt} 
{\small
       \begin{itemize}
        \item $(s_0,s,tr)\models\mathsf{wait\_in}(I,ch,\{\mathsf{d,v}\Rightarrow P\}) $  iff one of the following is satisfied: 
        \vspace{-5pt}
    \[\begin{array}{l}
    1.\, (s_0,s,tr')\models P|_{\mathsf{d}=0,\mathsf{v}=v}\wedge tr=\langle ch?,v\rangle^\chop tr'\\
    2.\, (s_0,s,tr')\models P|_{\mathsf{d}=d,\mathsf{v}=v} \wedge d>0 \wedge \overrightarrow{p}(0) = s_0\wedge\forall\,t\in [0,d].\,(s_0,t,\overrightarrow{p}(t))\models I\\
    \quad \wedge tr=\langle d,\overrightarrow{p},\{ch?\}\rangle^\chop\langle ch?,v\rangle^\chop tr'
    \end{array} \]
    \vspace{-10pt} 
        \item 
      $ (s_0,s,tr)\models\mathsf{wait\_outv}(I,ch,e,\{\mathsf{d}\Rightarrow P\}) $  iff one of the following is satisfied: 
      \vspace{-5pt} 
    \[\begin{array}{l}   
    1.\, (s_0,s,tr')\models P|_{\mathsf{d}=0}\wedge tr=\langle ch!,s_0(e)\rangle^\chop tr'\\
    2.\, (s_0,s,tr')\models P|_{\mathsf{d}=d} \wedge d>0 \wedge \overrightarrow{p}(0) = s_0\wedge\forall\,t\in [0,d].\,(s_0,t,\overrightarrow{p}(t))\models I\\
    \quad \wedge tr=\langle d,\overrightarrow{p},\{ch!\}\rangle^\chop\langle ch!,s_0(e)\rangle^\chop tr'
    \end{array} \]
    \vspace{-10pt} 
    \item $(s_0,s,tr)\models\mathsf{wait}(I,e,\{\mathsf{d}\Rightarrow P\})$  iff one of the following is satisfied: 
    \vspace{-5pt} 
    \[
     \begin{array}{l}
     1.\, (s_0,s,tr)\models P|_{\mathsf{d}=0}\wedge s_0(e)\le 0\\
    2.\, (s_0,s,tr')\models P|_{\mathsf{d}=s_0(e)} \wedge s_0(e)> 0 \wedge \overrightarrow{p}(0) = s_0\wedge\forall\,t\in [0,s_0(e)].\,(s_0,t,\overrightarrow{p}(t))\models I\\
    \quad \wedge tr=\langle s_0(e),\overrightarrow{p},\{\}\rangle^\chop tr'
    \end{array}\]
    \vspace{-10pt} 
    \item $(s_0,s,tr)\models\mathsf{interrupt}(I,e,\{\mathsf{d}\Rightarrow P\},cm)$  iff one of the following is satisfied: 
    \vspace{-5pt} 
    \[
    \begin{array}{l}
1.\,(s_0,s,tr)\models P|_{\mathsf{d}=0}\wedge s_0(e)\le 0\\
2.\,(s_0,s,tr')\models P|_{\mathsf{d}=s_0(e)}\wedge s_0(e)>0\wedge \overrightarrow{p}(0)=s_0 \wedge\forall\,t\in [0,s_0(e)].\,(s_0,t,\overrightarrow{p}(t))\models I\\
    \quad \wedge tr=\langle s_0(e),\overrightarrow{p},rdy(cm)\rangle^\chop tr'\\
3.\, (s_0,s,tr')\models P_i|_{\mathsf{d}=0,\mathsf{v}=v}\wedge cm[i]=(ch_i?,\{\mathsf{d,v}\Rightarrow P_i\})\wedge tr=\langle ch_i?,v\rangle^\chop tr'\\
4. \, (s_0,s,tr')\models P_i|_{\mathsf{d}=d,\mathsf{v}=v}\wedge cm[i]=(ch_i?,\{\mathsf{d,v}\Rightarrow P_i\})\wedge0<d\le s_0(e)\\
\quad\wedge \overrightarrow{p}(0)=s_0\wedge \forall\,t\in [0,d].\,(s_0,t,\overrightarrow{p}(t))\models I\wedge tr=\langle d,\overrightarrow{p},rdy(cm)\rangle^\chop\langle ch_i?,v\rangle^\chop tr'\\
5.\, (s_0,s,tr')\models P_i|_{\mathsf{d}=0}\wedge cm[i]=(ch_i!,h,\{\mathsf{d}\Rightarrow P_i\})\wedge tr=\langle ch_i!,h(0)\rangle^\chop tr'\\
6. \, (s_0,s,tr')\models P_i|_{\mathsf{d}=d}\wedge cm[i]=(ch_i!,h,\{\mathsf{d}\Rightarrow P_i\})\wedge0<d\le s_0(e)\\
\quad\wedge \overrightarrow{p}(0)=s_0\wedge \forall\,t\in [0,d].\,(s_0,t,\overrightarrow{p}(t))\models I\wedge tr=\langle d,\overrightarrow{p},rdy(cm)\rangle^\chop\langle ch_i!,h(d)\rangle^\chop tr'
\end{array}
\]
\item $(s_0,s,tr)\models\mathsf{interrupt_\infty}(I,cm)$  iff one of the following is satisfied: 
    \vspace{-5pt} 
    \[
    \begin{array}{l}
1.\, (s_0,s,tr')\models P_i|_{\mathsf{d}=0,\mathsf{v}=v}\wedge cm[i]=(ch_i?,\{\mathsf{d,v}\Rightarrow P_i\})\wedge tr=\langle ch_i?,v\rangle^\chop tr'\\
2. \, (s_0,s,tr')\models P_i|_{\mathsf{d}=d,\mathsf{v}=v}\wedge cm[i]=(ch_i?,\{\mathsf{d,v}\Rightarrow P_i\})\wedge0<d\\
\quad\wedge \overrightarrow{p}(0)=s_0\wedge \forall\,t\in [0,d].\,(s_0,t,\overrightarrow{p}(t))\models I\wedge tr=\langle d,\overrightarrow{p},rdy(cm)\rangle^\chop\langle ch_i?,v\rangle^\chop tr'\\
3.\, (s_0,s,tr')\models P_i|_{\mathsf{d}=0}\wedge cm[i]=(ch_i!,h,\{\mathsf{d}\Rightarrow P_i\})\wedge tr=\langle ch_i!,h(0)\rangle^\chop tr'\\
4. \, (s_0,s,tr')\models P_i|_{\mathsf{d}=d}\wedge cm[i]=(ch_i!,h,\{\mathsf{d}\Rightarrow P_i\})\wedge0<d\\
\quad\wedge \overrightarrow{p}(0)=s_0\wedge \forall\,t\in [0,d].\,(s_0,t,\overrightarrow{p}(t))\models I\wedge tr=\langle d,\overrightarrow{p},rdy(cm)\rangle^\chop\langle ch_i!,h(d)\rangle^\chop tr'
\end{array}
\]
\vspace{-10pt} 
    \end{itemize}
}
\vspace{-5pt} 
As defined by $\mathsf{wait\_in}$, the first case corresponds to communicating immediately, so the delay $\mathsf{d}$ is 0, the input value $\mathsf{v}$ can be any real number $v$ which can't be determined by itself. We use the notation $P|_{\mathsf{d}=0,\mathsf{v}=v}$ to represent the assertion obtained by replacing the appearance of $\mathsf{d}$ and $\mathsf{v}$ in $P$ with value 0 and $v$.
The second case corresponds to communicating after waiting for time $d>0$. The path taken by the state during waiting is given by $\overrightarrow{p}$, which satisfies the path condition $I$. $\mathsf{wait\_out}$ is defined similarly, but unlike the input case, the output value is determined by $e$ and the map $\{\mathsf{d}\Rightarrow P\}$ is only over the delay $\mathsf{d}$. 
For the  $\mathsf{wait}$ assertion, $e$ is a real expression specifying the wait time and the map in this assertion only has one argument over delay $\mathsf{d}$.  

The $\mathsf{interrupt}$ assertion covers two main categories of termination for the ODE. One occurs when the ODE evolves for zero or more time units and then terminates due to violating the domain. The other happens when the ODE, after evolving for zero or more time units, is interrupted by an input or output event.
In the definition of $\mathsf{interrupt}$ assertion, $e$ specifies the \emph{maximum} waiting time of the interrupt, $P$ specifies the remaining behavior if the waiting stops upon reaching the time bound $e$, $\cm$ specifies the list of communications that can happen at any time not exceeding $s_0(e)$. $\cm$ is given by a list of elements like $\langle ch_i?,\{\mathsf{d,v}\Rightarrow P_i\}\rangle$ or $\langle ch_i!,g,\{\mathsf{d}\Rightarrow P_i\}\rangle$, which specifies what happens after the corresponding interrupt is triggered, where $g$ is a function mapping from delay to the output value and $\rdy(\cm)$ denotes the ready set of communications in $\cm$.
There is an important special case: often we know the maximum waiting time may be infinite, for example when the domain of the ODE is true, the system can only execute the next command when a communication occurs. We denote this case by assertion $\mathsf{interrupt_\infty}(I,cm)$. 

Finally, we introduce the recursion assertion: 
\vspace{-5pt}
{\small
  \[
    \begin{array}{l}      (s_0,s,tr)\models\mathsf{Rec}\ R.\  P\bar{\vee} F(R) \mbox{ iff } (s_0,s,tr)\models P \mbox{ or } (s_0,s,tr)\models F(\mathsf{Rec}\ R.\  P\bar{\vee} F(R))
    \end{array}
    \vspace{-5pt}
    \]
}
\par\noindent
We can deduce that $(s_0,s,tr)\models\mathsf{Rec}\ R.\  P\bar{\vee} F(R)$
iff $\exists n. \, (s_0,s,tr)\models F^n(P)$ where $F^n\triangleq F(F^{n-1}(P))$ and $n$ is a natural number. 

\subsection{Specification}
In previous HHL~\cite{DBLP:journals/corr/abs-2303-15020}, the specification of a HCSP process $pc$ takes the form of Hoare triple $\{Pre\}\,pc\,\{Post\}$, where $Pre$ and $Post$ are predicates on state and trace. We use $(s,tr)\models Pre$ to denote that the state $s$ and the trace $tr$ satisfy the predicate $Pre$ ($Post$ is similar). 
Note that, an assertion $Q$ is a predicate over three elements: initial state $s_0$, current state $s$ and a trace $tr$, thus $Q(s_0)$ can be seen as a predicate on state and trace, e.g. $(s,tr)\models Q(s_0)\equiv(s_0,s,tr)\models Q$.
The validity of a Hoare triple is defined in terms of big-step  semantics as follows:
\vspace{-5pt}
{\small
\[
\begin{array}{l}
     \{Pre\}\,pc\,\{Post\} \triangleq  \\
     \qquad \forall s_1\ s_2\ tr\ tr'.\, (s_1,tr)\models Pre 
\longrightarrow (pc,s_1)\Rightarrow (s_2,tr')
\longrightarrow (s_2,tr^\chop tr')\models Post
\end{array}
\vspace{-5pt}
\]}
\par\noindent
In this paper, we utilize a new method of specification definition named $\mathsf{spec\_of}$ based on Hoare triples:
\vspace{-5pt}
{\small
\[
\mathsf{spec\_of}(pc,Q) \quad\triangleq\quad \forall s_0.\,\{s=s_0\land tr=\epsilon\}\,pc\,\{(s,tr)\models Q(s_0)\}
\vspace{-5pt}
\]}
\par\noindent
where the assertion $Q$ describes the relationship between the initial state $s_0$, the final state $s$ and the produced trace $tr$.
This specification means that if this process starts with a state $s_0$, then when the process terminates, the end state and the trace produced should meet the predicate $Q(s_0)$. 

\oomit{
In summary, our assertion language is to write the postcondition on state and trace in a fixed form of $R(s_0)$, where $s_0$ represents the initial state of the process and $R$ is an assertion constructed according to the syntax of assertions. The assertion $R$ describes the relationship between the initial state $s_0$, the final state $s$ and the produced trace $tr$.
}

Next, we give some useful characteristics and lemmas on predicates and assertions.

Given two predicates $G_1$ and $G_2$ , we define the entailment between $G_1$ and $G_2$ as:
\vspace{-5pt}
{\small
\[
G_1\Longrightarrow_a G_2 \quad\triangleq\quad\forall\  s\ tr.\, (s,tr)\models G_1 \longrightarrow (s,tr)\models G_2
\vspace{-5pt}
\]}

Obviously, this entailment relationship satisfies the transitivity and reflexivity.
There are some common entailment rules, for example introduction and elimination rules for conjunction or disjunction. Some special notes of entailment related to monotonicity and substitution of assertions are stated in the following.

The assertions $\mathsf{wait\_in}$, $\mathsf{wait\_outv}$, $\mathsf{wait}$, etc. all satisfy monotonicity rules on the initial state $s_0$, that reduce entailment relations among assertions to entailments on its components. For example, monotonicity of $\mathsf{wait\_in}$ take the following form:
\vspace{-5pt}
{\small
\[
\prftree{\forall d\ v.\, P_1|_{\mathsf{d}=d,\mathsf{v}=v}(s_0) \Longrightarrow_a P_2|_{\mathsf{d}=d,\mathsf{v}=v}(s_0)}
{\mathsf{wait\_in}(I,ch,\{\mathsf{d,v}\Rightarrow P_1\})(s_0)\Longrightarrow_a \mathsf{wait\_in}(I,ch,\{\mathsf{d,v}\Rightarrow P_2\})(s_0)}
\vspace{-5pt}
\]}
\par\noindent
This rule permits deducing entailment between two $\mathsf{wait\_in}$ assertions that differ only in the ensuing parameters. There are similar rules for $\mathsf{wait\_outv}$, $\mathsf{wait}$, $\mathsf{interrupt}$ and $\mathsf{interrupt_\infty}$.
By these rules, we can assert that all the functions from assertions to assertions constructed by the forms introduced satisfies monotonicity.

The commutativity with existential quantifier for assertions is like the following:
\vspace{-5pt}
{\small
\[
\mathsf{wait\_in}(I,ch,\{\mathsf{d,v}\Rightarrow \exists x.\,P\})(s_0) \Longrightarrow_a \exists x.\,\mathsf{wait\_in}(I, ch, \{\mathsf{d,v}\Rightarrow P\})(s_0)
\vspace{-5pt}
\]}
\par\noindent
Other forms of assertions in our logic have similar results. 
So far, both the monotonicity and commutativity conditions are proved to hold for the assertions defined at the beginning of this section. 
We proved in Isabelle that the $\mathsf{Rec}$ assertion is the least fixed point under the assumption that $F$ is monotonic  with respect to logical implication  and commutative with existential quantifier.

Besides, performing substitution $[x:=e]$ on assertions such as $\mathsf{wait\_in}$ can be reduced to performing the same operations on its components. For example, the entailment rule for $\mathsf{wait\_in}$ is:
\vspace{-5pt}
{\small
\[
\mathsf{wait\_in}(I,ch,\{\mathsf{d,v}\Rightarrow P\})[x := e](s_0) \Longrightarrow_a \mathsf{wait\_in}(I[x := e], ch, \{\mathsf{d,v}\Rightarrow P[x := e]\})(s_0)
\vspace{-5pt}
\]}

\section{Inference Rules for Sequential HCSP}
\label{SequentialHCSP}
In this section, we introduce the inference rules for generating assertions of sequential HCSP processes.
For each sequential HCSP construct, we define the rule for it where it is followed by a subsequent process $c$. This is because different processes can have varying effects on the traces of the sequentially composed $c$. Notably,  the rules for the constructs alone can be derived by substituting $c$ with $\pskip$ and applying the skip rule.  

For skip, assignment, input, output, wait and if commands, we have following rules:
\vspace{-5pt}
{\small
\[
\begin{array}{c}
     \prftree{\mathsf{spec\_of}(\pskip, \mathsf{init})} \vspace{1mm} 
\quad
\prftree{\mathsf{spec\_of}(c,Q)}
{\mathsf{spec\_of}(\pskip; c, Q)}
\quad
\prftree{\mathsf{spec\_of}(c,Q)}
{\mathsf{spec\_of}(x := e; c, Q[x := e])}
\\
\prftree{\mathsf{spec\_of}(c_1;c, P)}{\mathsf{spec\_of}(c_2;c, Q)}
{\mathsf{spec\_of}(\IFE{B}{c_1}{c_2}; c, (\uparrow(B)\bar{\land} P) \bar{\lor} (\uparrow(\neg B)\bar{\land} Q))}
\\
\prftree{\mathsf{spec\_of}(c, Q)}
{\mathsf{spec\_of}(ch?x; c, \mathsf{wait\_in}(\mathsf{id}, ch, \{\mathsf{d,v}\Rightarrow Q[x := \mathsf{v}]\}))}\\
\prftree{\mathsf{spec\_of}(c, Q)}
{\mathsf{spec\_of}(ch!e; c, \mathsf{wait\_outv}(\mathsf{id}, ch, e, \{\mathsf{d}\Rightarrow Q\}))}\\
\prftree{\mathsf{spec\_of}(c, Q)}
{\mathsf{spec\_of}(\pwait \ e; c, \mathsf{wait}(\mathsf{id}, e, \{\mathsf{d}\Rightarrow Q\}))}
\end{array}
\]}

For the nondeterministic repetition command, we have the following rule:
\vspace{-5pt}
{\small
\[
\begin{array}{c}
\prftree{\mathsf{spec\_of}(c',P)}
{\forall\ cc\ Q.\ \mathsf{spec\_of}(cc,Q) \longrightarrow \mathsf{spec\_of}(c;cc,F(Q))}
{\mathsf{spec\_of}(c^*;c',\mathsf{Rec}\ R.\  P\bar{\lor} F(R))}
\end{array}
\vspace{-5pt}
\]}
\par\noindent
In this rule, $P$ represents the assertion of proceeding directly to subsequent processes without executing the loop and $F$ represents
the change in assertion resulting from executing once loop. This recursion assertion can be seen as the loop invariant of repetition.

We now state the rules for continuous evolution. If the (unique) solution to the ODE is known, the predicate $\mathsf{paramODEsol}(\overrightarrow{\dot{x}}=\overrightarrow{e}, B, f, e)$ is introduced: 
$\overrightarrow{\dot{x}}=\overrightarrow{e}$ is an equation between variables and their derivative expressions;
 $B$ is a predicate on the state, specifying the open boundary condition;
 $f(\overrightarrow{x},t)$ is the solution of $\overrightarrow{\dot{x}}=\overrightarrow{e}$ at time $t$;
  $e$ maps the starting state to the length of time for the unique solution of the ODE reaching the boundary.
 We can then state the inference rule for the continuous evolution as follows:
 \vspace{-5pt}
{\small
\[
\begin{array}{c}
\prftree{\mathsf{paramODEsol}(\overrightarrow{\dot{x}}=\overrightarrow{e}, B, f, e)}{\mathsf{lipschitz}(\overrightarrow{\dot{x}}=\overrightarrow{e})}{\mathsf{spec\_of}(c, Q)}
{\mathsf{spec\_of}(\langle \overrightarrow{\dot{x}}=\overrightarrow{e}\& B\rangle; c, \mathsf{wait}(
\overrightarrow{x}\rightarrowtail f(\overrightarrow{x},t), e, \{\mathsf{d} \Rightarrow Q[\overrightarrow{x}:=f(s_0(\overrightarrow{x}),\mathsf{d})] \}))}
\end{array}
\vspace{-5pt}
\]}
\par\noindent
The meaning of this rule is as follows. Suppose $\overrightarrow{\dot{x}}=\overrightarrow{e}$ with boundary condition $B$ has solution $f$ with time given by $e$ (both functions of $s_0$) and the  lipschitz predicate ensures that there is a unique solution
to this ODE, then the specification of $\langle \overrightarrow{\dot{x}}=\overrightarrow{e}\& B\rangle; c$ first evolves along the path $\overrightarrow{p}(t)=s0[\overrightarrow{x}\mapsto f(s_0(\overrightarrow{x}),t)]$ for time $s_0(e)$, then followed by the behavior of $c$ as specified by $Q$ starting from the updated state $s_0[\overrightarrow{x}:=f(s_0(\overrightarrow{x}),\mathsf{d})]$.

Next, we show how to use differential invariants to reason about  continuous evolution. We define predicate $\mathsf{paramODEInv}(\overrightarrow{\dot{x}}=\overrightarrow{e},inv,pp)$, meaning that if the starting state of ODE satisfies the condition $pp$, then all the states along the ODE $\overrightarrow{\dot{x}}=\overrightarrow{e}$ satisfy the invariant $inv$.
Before applying this rule, we should have $inv$ and corresponding differential methods provided. The predicate is verified using the technology introduced in
\cite{DBLP:books/sp/Platzer18,DBLP:conf/fm/ShengBZ23}. 
\vspace{-5pt}
{\small
\[
\prftree{\mathsf{paramODEInv}(\overrightarrow{\dot{x}}=\overrightarrow{e}, B, inv, pp)}{\mathsf{lipschitz}(\overrightarrow{\dot{x}}=\overrightarrow{e})}{\mathsf{spec\_of}(c,Q)}
{
\begin{array}{c}
     \mathsf{spec\_of}(\langle \overrightarrow{\dot{x}}=\overrightarrow{e}\& B\rangle;c, (\uparrow(\neg B)\bar{\wedge} Q)\bar{\vee}\uparrow(\neg pp\wedge B)\bar{\vee}\\
     \exists\,T\,\overrightarrow{nx}.\,(\uparrow (pp\wedge B)\bar{\wedge}\mathsf{wait}(
    inv, T, \{d \Rightarrow (\uparrow(inv\wedge bound(B)) \bar{\wedge} Q)[\overrightarrow{x}:=\overrightarrow{nx}] \})))
\end{array}
}
\vspace{-5pt}
\]}
\par\noindent
This rule includes three cases via disjunction: (1) If the boundary is violated at the beginning, then the ODE terminates at once and satisfies the specification of $c$. (2) The second case is when the condition $pp$ does not hold. Although we do not desire this situation to arise, it must be included to ensure the correctness of the specification. We expect $\neg pp$ to conflict with other conditions in the subsequent verification and counteract this case, indicating that this case will not happen. (3) The last case states that it will stop at some state satisfying both the invariant and the boundary of $B$. (During implementation, we will introduce new variables $T$ and $\overrightarrow{nx}$ to avoid Existential quantifier.)

The inference rules for interrupt command can be seen as the combination of rules for ODE, input and output. We put them in Appendix \ref{App:seq} for the sake of brevity.
Below we give an example to illustrate how to generate the specifications of sequential HCSP processes by  applying these rules. 

\begin{example}
This example  illustrates handling of delay and communication events. 
\vspace{-5pt}
{\small
\[
c\triangleq ch_2?x; \pwait\ 1; ch_1!x
\vspace{-5pt}
\]
}
\par\noindent
The specification of $c$ is generated by the following steps:
{\small
\[
\begin{array}{cl}
      1: &\mathsf{spec\_of}(ch_1!x,\mathsf{wait\_outv}(\mathsf{id}, ch_1, x, \{\mathsf{d1}\Rightarrow \mathsf{init}\})) \\
     2: &\mathsf{spec\_of}(\pwait(1); ch_1!x,\mathsf{wait}(\mathsf{id},1,\{\mathsf{d2}\Rightarrow \mathsf{wait\_outv}(\mathsf{id}, ch_1, x, \{\mathsf{d1}\Rightarrow \mathsf{init}\})\})) \\
      3: &\mathsf{spec\_of}(ch_2?x;\pwait(1); ch_1!x,\mathsf{wait\_in}(\mathsf{id}, ch_2, \{\mathsf{d3,v3}\Rightarrow \\
     &\quad\mathsf{wait}(\mathsf{id},1,\{\mathsf{d2}\Rightarrow \mathsf{wait\_outv}(\mathsf{id}, ch_1, x, \{\mathsf{d1}\Rightarrow \mathsf{init}\})\})[x := \mathsf{v3}]\}))\\
     4: &\mathsf{spec\_of}(ch_2?x;\pwait(1); ch_1!x,\mathsf{wait\_in}(\mathsf{id}, ch_2, \{\mathsf{d3,v3}\Rightarrow \\
     &\quad\mathsf{wait}(\mathsf{id} [ x := \mathsf{v3} ],1,\{\mathsf{d2}\Rightarrow  \mathsf{wait\_outv}(\mathsf{id}[x:=\mathsf{v3}], ch_1, \mathsf{v3}, \{\mathsf{d1}\Rightarrow \mathsf{init}\})\})\}))
\end{array}
\]}

At Step 4, we obtain the final specification of $c$, which can be understood as follows: Starting from state $s_0$, first waits for input along channel $ch_2$, after receiving input value $\mathsf{v3}$ at time $\mathsf{d3}$, then waits for time 1 with state $s_0[x:=\mathsf{v3}]$, then waits for output along channel $ch_1$ with state $s_0[x:=\mathsf{v3}]$, that occurs at time $\mathsf{d1}$. The output value is $\mathsf{v3}$, and the final state after output is $s_0[x:=\mathsf{v3}]$.

\end{example}



\section{Inference Rules for Parallel HCSP}
\label{ParallelHCSP}
In this section, we introduce the inference rules for constructing assertions of parallel processes by synchronization.  
In order to handle parallel processes, we define operator $\mathsf{sync}(chs,P_1,P_2)$ denoting the synchronization if given two assertions $P_1$ and $P_2$ for two processes and the set of common channels $chs$ through which communications occur between them:
\vspace{-5pt}
{\small
\[
\begin{array}{c}
     (s_0, s, tr)\models \mathsf{sync}(chs,P_1,P_2)\text{ iff }  \exists\, s_{01}\, s_{02}\, s_{1}\,s_{2}\,tr_1\,tr_2. s_0=s_{01}\uplus s_{02} \wedge s=s_{1}\uplus s_{2} \wedge\\
     (s_{01},s_{1},tr_1)\models P_1\wedge
(s_{02},s_{2},tr_2)\models P_2\wedge
tr_1\|_{chs}tr_2 \Downarrow tr 
\end{array}
\vspace{-5pt}
\]}
\par\noindent
By the above definition of $\mathsf{sync}$, we can easily obtain the following conclusion:
\vspace{-5pt}
{\small
\[
\prftree{\mathsf{spec\_of}(c_1,P_1)}{\mathsf{spec\_of}(c_2,P_2)}
{\mathsf{spec\_of}(c_1\|_{chs}c_2, \mathsf{sync}(chs,P_1,P_2))}
\vspace{-5pt}
\]
}
\par\noindent
However, we can't intuitively derive valid information from the definition of this operator.
Our objective is to find an assertion $Q$ within our assertion language that can replace $\mathsf{sync}(chs,P_1,P_2)$, ensuring that $Q$ is logically implied by  $\mathsf{sync}(chs,P_1,P_2)$ and thus satisfies the above specification. 
We conclude this motivation to reach the following inference rule for parallel composition: 
\vspace{-5pt}
{\small
\[
\prftree{\mathsf{spec\_of}(c_1,P_1)}{\mathsf{spec\_of}(c_2,P_2)}
{\forall\ s_0.\, \mathsf{sync}(chs,P_1,P_2)(s_0) \Longrightarrow_a Q(s_0)}
{\mathsf{spec\_of}(c_1\|_{chs}c_2, Q)}
\vspace{-5pt}
\]
}
\par\noindent
We hope that $Q$ reserves the whole behaviour of parallel process to facilitate verification of the system in subsequent steps.  
For example, the trivial $\mathsf{true}$ is always satisfactory, but we can't get any valid information from it.
Thus, our proof system contains a set of inference rules for reasoning about the parallel synchronization of assertions in the form of $\mathsf{sync}(chs,P,Q)(s_0)\Longrightarrow_a Q(s_0)$.

By repeatedly using synchronization rules (as well as monotonicity rules and other entailments among assertions), we can gradually reduce an assertion headed by $\mathsf{sync}$ into one without $\mathsf{sync}$ operators. For the sake of brevity, we select a representative case to illustrate the synchronization rules. The following rule states that, when the channels of two sides match, the communication occurs immediately, determining the time variable $\mathsf{d}$ with 0 and the value variable $\mathsf{v}$ with $e(s_0)$, and then the procedure of synchronization continues to the tail assertions $P_1$ and $P_2$. 
\vspace{-5pt}
{\small
\[
\prftree[r]{InOut1}{ch_1\in chs}{ch_2\in chs}{ch_1=ch_2}{
\begin{array}{c}
     \mathsf{sync}(chs,\mathsf{wait\_in}(I_1,ch_1,\{\mathsf{d,v}\Rightarrow P_1\}),\mathsf{wait\_outv}(I_2,ch_2,e,\{\mathsf{d}\Rightarrow P_2\}))(s_0)\Longrightarrow_a  \\
     \mathsf{sync}(chs,P_1|_{\mathsf{d}=0,\mathsf{v}=s_0(e)},P_2|_{\mathsf{d}=0})(s_0)
\end{array}
}
\vspace{-5pt}
\]}
\par\noindent
We present other rules in Appendix~\ref{App:par} and explain their intuitive meanings. The soundness of these rules have been formally proven by combining the definition of operator $\mathsf{sync}$ and the trace synchronization relation as introduced in Sect.~\ref{sec:hcsp}.

\begin{example}
This example demonstrates the handling of communication synchronization and loop. It repeatedly sends the same value $x$ from the left to the right, with $z$ received on the right, and then sends $z+1$ back from the right to the left. 
\vspace{-5pt}
{\small
\[
c_1\triangleq (ch_1!x;ch_2?y)^*\qquad
c_2\triangleq (ch_1?z;ch_2!(z+1))^*
\vspace{-5pt}
\]}
\par\noindent
By applying the rules for input, output, sequential composition and repetition, we can derive $ 
\mathsf{spec\_of}(c_1,P_1)$ and $\mathsf{spec\_of}(c_2,P_2)
$
with
\vspace{-5pt}
{\small
\[
\begin{array}{l}
    P_1\triangleq \mathsf{Rec}\ R_1.\ \mathsf{init}\bar{\lor} \mathsf{wait\_outv}(\mathsf{id},ch_1,x,\{\mathsf{d_1}\Rightarrow \\
    \qquad
    \mathsf{wait\_in}(\mathsf{id},ch_2,\{\mathsf{d_2,v_2}\Rightarrow R_1[x:=\mathsf{v_2}]\})\})\\
    P_2\triangleq \mathsf{Rec}\ R_2.\ \mathsf{init}\bar{\lor} \mathsf{wait\_in}(\mathsf{id},ch_1,\{\mathsf{d_1,v_1}\Rightarrow \\
    \qquad
    \mathsf{wait\_outv}(\mathsf{id[z:=\mathsf{v_1}]},ch_2,\mathsf{v_1}+1,\{\mathsf{d_2}\Rightarrow R_2[z:=\mathsf{v_1}]\})\}
\end{array}
\vspace{-5pt}
\]
}
\par\noindent
According to the rule for synchronization of two recursion assertions, we can derive
\vspace{-5pt}
{\small
\[
\mathsf{sync}(\{ch_1,ch_2\},P_1,P_2)(s_0)\Longrightarrow_a
\mathsf{Rec}\ R.\  \mathsf{init}\bar{\lor}R[z:=s_0(x)][y:=s_0(x)+1](s_0)
\vspace{-5pt}
\]
}
\end{example}
As indicated by the final specification, the internal communications over the common channel set $\{ch_1,ch_2\}$ are hidden and unobservable. The effect of the parallel composition of $c_1$ and $c_2$ is to repeatedly assign $z$ the value of $x$ and assign $y$ the value of $x+1$ to their joint state $s_0$, iterated any number of times.

\section{Property Verification}
\label{Propertyv}
Till now, we have introduced the inference rules to generate the assertion $Q$ satisfying $\mathsf{spec\_of}(pc,Q)$, for either sequential or parallel processes $pc$. As defined by the semantics of assertions in Sect.~\ref{Assertions},  $Q$ captures the trace execution history of $pc$ over time up to the termination of $pc$. However, deriving the properties of process $pc$ (related to variables) during execution directly from assertion $Q$ is far from intuitive.
In this section, we present how to verify  properties of a process in a fixed form
 of $(s,tr)\models Post \triangleq q_1(s)\land \mathsf{trI}(tr,q_2) $ where  $s$ and $tr$ represent the final state and trace at termination, $q_1$ and $q_2$ are boolean expressions on state, and 
\vspace{-4pt}
{\small
\[
\mathsf{trI}(tr,q)\triangleq \forall i. \, tr[i]=\langle d,\overrightarrow{p},rdy\rangle \longrightarrow (\forall t\in [0,d]. \, q(\overrightarrow{p}(t)))
\vspace{-4pt}
\]}
\par\noindent
Intuitively speaking, $Post$ holds for final state $s$ and trace $tr$, iff $q_1$ holds for the final state $s$, and $q_2$ holds for each continuous state in $tr$, i.e. it holds almost everywhere during the whole execution of $pc$ (except for some discrete events). 
In the following, we will call $q_1$ and $q_2$ postcondition and trace invariant respectively.
Together with the definition of specification, we conclude the following inference rule:
\vspace{-4pt}
{\small
\[
\prftree{\forall \,s_0\,s\,tr. \,p(s_0)\longrightarrow (s_0,s,tr)\models Q\longrightarrow (s,tr)\models Post}{\mathsf{spec\_of}(pc,Q)}{\{Pre\}\,pc\,\{Post\}}
\vspace{-4pt}
\]
}
\par\noindent
where $(s,tr)\models Pre\triangleq p(s)\wedge tr=\epsilon$ which represents that the process $pc$ starts from an initial state satisfying precondition $p$ and an empty trace.
Next, we present how to derive the first antecedent of the above rule for different forms of assertions.
We only consider closed processes $pc$ for which all communications are internal, thus no communications are contained in $Q$ any more as all internal communications are reduced during synchronization, as shown in rule InOut1. 

For the $\mathsf{init}$ assertion, we have :
\vspace{-5pt}
{\small
\[
\prftree{\forall s.\,p(s)\longrightarrow q_1(s)}{p(s_0)\longrightarrow (s_0,s,tr)\models \mathsf{init}\longrightarrow q_1(s)\land \mathsf{trI}(tr,q_2)}
\vspace{-5pt}
\]}
\par\noindent
since $\mathsf{init}(s_0,s,tr)$ implies $s=s_0$ and $tr=\epsilon$.

For the $\mathsf{wait}$ assertion, we have
\vspace{-5pt}
{\small
\[
\prftree{
\begin{array}{c}
     p(s_0)\land s_0(e)>0 \land t\ge 0\land t\le s_0(e) \longrightarrow  (s_0,t,s)\models I \longrightarrow q_2(s)\\
     p(s_0)\land s_0(e)>0\longrightarrow (s_0,s,tr)\models P|_{\mathsf{d}=s_0(e)}\longrightarrow q_1(s)\land \mathsf{trI}(tr,q_2)  \\
     p(s_0)\land s_0(e)\le0\longrightarrow (s_0,s,tr)\models P|_{\mathsf{d}=0}\longrightarrow q_1(s)\land \mathsf{trI}(tr,q_2)  
\end{array}
}{p(s_0)\longrightarrow(s_0,s,tr)\models\mathsf{wait}(I,e,\{\mathsf{d}\Rightarrow P\})\longrightarrow q_1(s)\land \mathsf{trI}(tr,q_2)}
\vspace{-5pt}
\]}
\par\noindent
where the wait time is evaluated (either positive or not) to determine the remaining part and check the trace invariant from the path condition.

For pure assertion, we have:
\vspace{-5pt}
{\small
\[
\prftree{p(s_0)\land b(s_0)\longrightarrow (s_0,s,tr)\models Q\longrightarrow (s,tr)\models Post}{p(s_0)\longrightarrow (s_0,s,tr)\models(\uparrow b\bar{\land} Q)\longrightarrow (s,tr)\models Post}
\vspace{-5pt}
\]}

For substitution, we have:
\vspace{-5pt}
{\small
\[
\prftree{\forall s_0\,s\,tr.\, (\exists v.\, p[v/x]\land x=e[v/x])(s_0)\longrightarrow (s_0,s,tr)\models Q\longrightarrow (s,tr)\models Post}{p(s_0)\longrightarrow (s_0,s,tr)\models Q[x:=e]\longrightarrow (s,tr)\models Post}
\vspace{-5pt}
\]}
\par\noindent
As shown in this rule, we change the initial state from $s_0$ to $s_0[x\mapsto e]$, thus the precondition $p$ needs to be rewritten on the new state while maintaining the equivalence.

For disjunction, we have:
\vspace{-5pt}
{\small
\[
\prftree{
\begin{array}{c}
     p(s_0)\longrightarrow (s_0,s,tr)\models Q_1\longrightarrow (s,tr)\models Post  \\
     p(s_0)\longrightarrow (s_0,s,tr)\models Q_2\longrightarrow (s,tr)\models Post 
\end{array}
}
{p(s_0)\longrightarrow (s_0,s,tr)\models Q_1\bar{\vee} Q_2\longrightarrow (s,tr)\models Post}
\vspace{-5pt}
\]}

For recursion assertion, we have: 
\vspace{-5pt}
{\small
\[
\prftree{
\begin{array}{c}
     \forall s.\,p(s)\longrightarrow loop(s) \\ 
     \forall s_0\,s\,tr.\,loop(s_0)\longrightarrow (s_0,s,tr)\models P\longrightarrow (s,tr)\models Post\\
     \forall Q\,s_0\,s\,tr.\,(\forall s_0\,s\,tr.\,loop(s_0)\longrightarrow (s_0,s,tr)\models Q\longrightarrow (s,tr)\models Post)\\
     \longrightarrow
     loop(s_0)\longrightarrow (s_0,s,tr)\models F(Q)\longrightarrow (s,tr)\models Post 
\end{array}
}{p(s_0)\longrightarrow(s_0,s,tr)\models(\mathsf{Rec}\ R.\ P\bar{\lor} F(R))\longrightarrow (s,tr)\models Post}
\vspace{-5pt}
\]}
\par\noindent
In this rule we need to provide a loop invariant $loop$ and prove three conditions for $loop$ to be an invariant. The first two conditions states the precondition implies the loop invariant and the base assertion $P$ implies postcondition under the invariant. The intuitive meaning of the last one is that,
for any assertion $Q$, $F(Q)$ satisfying property $Post$ under loop invariant $loop$ can be deduced from that $Q$ satisfying  property $Post$ under $loop$. From this condition, we can extend the property to the general recursion $\mathsf{Rec}\ R.\ P\bar{\lor} F(R)$.
Since once loop means once $F$ applied to the assertion $P$, if we can prove $F(Q)$ satisfying the property $Post$ from any $Q$ have already meets it, then we can extend to $F^n(P)$ for any nature number $n$ of the loop times.

We demonstrate the usage of these rules by the following example involving delay and loop.
\begin{example}
\label{example_prop}
\vspace{-5pt}
{\small
\[
c\triangleq(\pwait\ 1;x:=x+1)^*
\vspace{-5pt}
\]}
\par\noindent
For process $c$, it's easy to find that if the initial state $s_0$ satisfies $x=1$, then $x>0$ will hold for the  final state at termination and also for each continuous state during the execution. This property can be described in Hoare triples as:
\vspace{-5pt}
{\small
\[
\{p(s) \land tr=\epsilon\}\,c\,\{q_1(s)\land \mathsf{trI}(tr,q_2)\}
\vspace{-5pt}
\]}
\par\noindent
where we define $p\triangleq x=1$, $q_1\triangleq x>0$ and $q_2\triangleq x>0$. To prove this triple, we apply the main inference rule resulting in two premises.
\vspace{-5pt}
{\small
\[
\mathsf{spec\_of}(c,\mathsf{Rec}\ R.\ \mathsf{init}\bar{\lor} \mathsf{wait}(\mathsf{id},1,\{\mathsf{d}\Rightarrow R[x:=x+1]\}))
\vspace{-5pt}
\]
}
\par\noindent
which can be derived by the sequential inference rules in Sect.~\ref{SequentialHCSP}, and
\vspace{-5pt}
{\small
\[
p(s_0)\longrightarrow(s_0,s,tr)\models(\mathsf{Rec}\ R.\ \mathsf{init}\bar{\lor} \mathsf{wait}(\mathsf{id},1,\{\mathsf{d}\Rightarrow R[x:=x+1]\}))\longrightarrow q_1(s)\land \mathsf{trI}(tr,q_2)
\vspace{-5pt}
\]
}
\par\noindent
which can be derived by rules in this section according to the structures of assertions by providing the loop invariant $loop\triangleq x>0$. The detailed proof is shown in Appendix~\ref{App:prop}. 
\end{example}

\section{Implementation and Case Study}
\label{sec:case}
In this section, we present the implementation of HHLPar and demonstrate its application through a case study. We formalize the underlying logic and establish its soundness using Isabelle/HOL, thereby ensuring the  correctness of the proof system. In addition to providing a correctness guarantee for the HHL logic, the Isabelle implementation also enables the interactive verification of HCSP by applying the appropriate inference rules. HHLPar is built on this logic and aims to enhance the automation of proof procedures. 



\subsection{HHLPar in Python}

We introduce HHLPar from two aspects: the overall structure, and the main implementation issues in Python. 

\subsubsection{HHLPar in a Nutshell}

\begin{algorithm}[t]
  \caption{Main algorithm of HHLPar}
  \label{alg:main}
  \begin{algorithmic}[1]
  \Require precondition $p$, HCSP process $pc$, postcondition $q_1$, trace invariant $q_2$, 
         additional: ode invariants $ode\_inv$, loop invariants $loop\_inv$
  \Ensure Success/Fail
    \Function{Assertion}{$pc,ode\_inv$}
        \If{$pc = pc_1\|_{chs}pc_2$}
            \State $P_1\gets\textsc{Assertion}(pc_1,ode\_inv)$
            \State $P_2\gets\textsc{Assertion}(pc_2,ode\_inv)$
            \State \Return $\textsc{ParSyn}(chs,P_1,P_2)$
        \Else 
            \State \Return $\textsc{SeqGen}(pc,ode\_inv)$
        \EndIf
    \EndFunction
    \State $P\gets \textsc{Assertion}(pc,ode\_inv)$
    \If{$\textsc{ProVer}(P,p,q_1,q_2,loop\_inv)$}
        \State\Return Success
    \Else
        \State\Return Fail
    \EndIf
  \end{algorithmic}
\end{algorithm}

The main algorithm of the HHLPar tool is illustrated in Alg.~\ref{alg:main}. The tool accepts as input: a precondition, a HCSP process to be verified, a postcondition and a trace invariant,  as well as additional invariants for ODEs and loops, if they are present. The verification process is carried out through three main steps: Sequential Generation, Parallel Synchronization, and Property Verification. The first step processes the sequential components of $pc$ and generates their assertions, and then the second step generates the assertion of $pc$ through synchronization of sequential ones. 
After these two steps, an assertion $Q$ satisfying $\mathsf{spec\_of}(pc,Q)$ will be obtained. The last step verifies whether postcondition and trace invariant hold for given precondition, with a result returned. 

\subsubsection{Implementation  in Python}

HHLPar implement the following three functionalities in the algorithm. 
\vspace{-10pt}
\paragraph{Sequential Generation}
We implemented the function for generating assertions of sequential HCSP satisfying the specification. 
When dealing with ODEs, if the differential invariants are not provided, this function  will invoke the Wolfram Engine to compute solutions in symbolic form and compute the maximum  waiting time based on constraint. For the sake of expressiveness and convenience, we choose to create a fresh time variable representing the length of this duration and record the constraints of this time variable in a boolean expression. For example, $\langle \dot{x}=1\& x<5\rangle$ corresponds to $\uparrow(t_1=5-x)\bar{\wedge}\mathsf{wait}(x\rightarrowtail x+t,t_1,\{\mathsf{d}\Rightarrow\mathsf{init}[x:=x+\mathsf{d}]\})$. If differential invariants are provided, this function will check whether the invariants are correct.

\vspace{-10pt}
\paragraph{Parallel Synchronization}
We implemented the synchronization function which accepting two assertions and the communication channel set and then producing the parallel assertion.
Note that variables in different processes are independent and cannot be shared in HCSP. Consequently, when same variable names occur in parallel processes and subsequently in their specifications, we consider them different. Therefore, before synchronization of assertions,  we assign process names to different  parallel processes and their corresponding assertions in the implementation. 
\vspace{-10pt}
\paragraph{Property Verification}
We implemented the verifying function which takes an assertion (the result of the previous steps), three boolean expressions representing the precondition on the initial state $s_0$, the postcondition on the final state $s$ and the trace invariant on the trace $tr$ separately and additional loop invariants as inputs. 
When applying the rules, the expression on initial state $s_0$ will be constantly updated.
When the assertion is a recursion, we need to prove that the loop invariant is maintained over each loop iteration.
This function will invoke the Wolfram Engine to check all the logical formulas in premises. If all of them are valid, the algorithm will stop successfully, indicating that this property is indeed satisfied with respect to the assertion and precondition, and in consequence it holds for the process being verified with the given Hoare triples.

\subsection{Case Study}
We experimented with a series of examples to test HHLPar across various situations. In this section, we illustrate its  ability  to handle simple branches in bulk through one case study, demonstrating how HHLPar can effectively verify processes involving ODEs, interrupts, communications, repetition and parallel composition involved.

The simplified case study of a cruise
control system (CCS) is taken from~\cite{tcs/XuWZJTZ22}, for which the verification was performed via interactive theorem proving. 
Compared to ~\cite{Xu2024}, we have implemented the algorithm for assertions to prove final properties of the process, and the whole procedure of verification is automated.
The model of the CCS comprises two parts: a controller (Control) and a physical plant (Plant). 
The Plant process models the vehicle's movement, continuously evolving along a given ODE. The evolution is periodically interrupted by the transmission of velocity $v$ and position $p$ to the Control, followed by the reception of updated acceleration $a$.
\vspace{-5pt}
{\small
\[
\textit{Plant} \triangleq \textit{ch1}!v;\textit{ch2}!p;(\textit{ch3}?a;\langle\dot{p}=v, \dot{v}=a \&\mathsf{true}\propto\pskip\rangle \unrhd \talloblong[{\textit{ch1}!v}\rightarrow {\textit{ch2}!p}])^* 
\vspace{-5pt}
\] 
}
\par\noindent
The Control process computes and sends the appropriate vehicle acceleration, determined by the received velocity and position, with respect to a period $T$.
\vspace{-5pt}
{\small
\[
\begin{array}{ll}
     &  \textit{Control} \triangleq \textit{ch1}?v; \textit{ch2}?p;(pp:=p+v\cdot T+\frac{1}{2}\cdot da\cdot T^2;vv:=v+da\cdot T;\\
     & \qquad\qquad\qquad 
     (\textrm{if}\ 2\cdot am\cdot(op-pp)\ge vm^2\ \textrm{then}\ vlm:=vm^2 \ \textrm{else}\\
     & \qquad\qquad\qquad\qquad 
     \textrm{if}\ op-pp>0\ \textrm{then}\ vlm:=2\cdot am\cdot(op-pp)\ \textrm{else}\ vlm:=0);\\
     & \qquad\qquad\qquad 
     (\textrm{if}\ vv\le 0\|vv^2\le vlm\ \textrm{then}\ a:=da\ \textrm{else}\ (pp:=p+v\cdot T;\\
     & \qquad\qquad\qquad\qquad
     (\textrm{if}\ 2\cdot am\cdot(op-pp)\ge vm^2\ \textrm{then}\ vlm:=vm^2 \ \textrm{else}
     \\
     & \qquad\qquad\qquad\qquad\qquad 
     \textrm{if}\ op-pp>0\ \textrm{then}\ vlm:=2\cdot am\cdot(op-pp)\ \textrm{else}\ vlm:=0);\\
     & \qquad\qquad\qquad\qquad 
     \textrm{if}\ v\le 0\|v^2\le vlm\ \textrm{then}\ a:=0\ \textrm{else}\ a:=-am));\\
     & \qquad\qquad\qquad
     \textit{ch3}!a;\pwait \ T; \textit{ch1}?v; \textit{ch2}?p)^*
\end{array}
\vspace{-5pt}
\] }
\par\noindent
where constants $T$, $op$, $ad$, $am$ represent the time period, the position of obstacle, the fixed acceleration during speeding up and deceleration separately, and the variable $vlm$ is the upper limit of velocity based on the concept of Maximum Protection Curve. 

In this case, the parallel process $Plant\|_{ch1,ch2,ch3}Control$ is provided to the tool HHLPar. The tool automatically gives \textit{Plant} (and \textit{Control}) and all the variables appearing in them a prefix name $A$ (and $B$) and the loop invariant $inv$ are provided below:
\vspace{-5pt}
{\small
\[
\begin{array}{cc}
     &  BT>0 
\land Bam>0 
\land Bda>0 
\land Bvm>0 
\land Ap\le Bop
\land Av=Bv
\land Ap=Bp
\\
     & \land \ ((2\cdot Bam\cdot(Bop-Ap)\ge Bvm^2 \land Av\le Bvm)\ \lor\\
     &
(2\cdot Bam\cdot(Bop-Ap)<Bvm^2 \land (Av\le0\lor Av^2\le2\cdot Bam\cdot(Bop-Ap))))
\end{array}
\vspace{-5pt}
\] 
}
\par\noindent
under the following provided precondition, denoted by $\textit{Init}$:
\vspace{-5pt}
{\small
\[
\begin{array}{cc}
     &  BT>0 
\land Bam>0 
\land Bda>0 
\land Bvm>0 
\land Ap\le Bop
\\
     & \land \ ((2\cdot Bam\cdot(Bop-Ap)\ge Bvm^2 \land Av\le Bvm)\ \lor\\
     &
(2\cdot Bam\cdot(Bop-Ap)<Bvm^2 \land (Av\le0\lor Av^2\le2\cdot Bam\cdot(Bop-Ap))))
\end{array}
\vspace{-5pt}
\]
}
\par\noindent
indicating the requirements on constants and that the initial position does not exceed the obstacle and the initial velocity is within the MPC, and $Ap \le Bop$ provided as both the postcondition and trace invariant, denoted by $\textit{Safe}$, HHLPar finally returns "pass".
This indicates that the following specification is proved:
\vspace{-5pt}
{\small
\[\{\textit{Init}(s) \land tr=\epsilon\}\,Plant\|_{ch1,ch2,ch3}Control\,\{\textit{Safe}(s)\land\mathsf{trI}(tr,\textit{Safe})\}
\vspace{-5pt}
\]
}

\section{Conclusion}
We presented HHLPar, an automated theorem prover for verifying parallel HCSP processes, which cover basic ingredients of hybrid and cyber-physical systems including discrete control, continuous dynamics, communication, interrupts and parallel composition. HHLPar implements a Hybrid Hoare Logic, that is composed of a set of inference rules for reasoning about sequential HCSP processes and a set of inference rules for reasoning about parallel HCSP processes, with the help of specialized assertions and their synchronization. HHLPar provides both guarantee to soundness from the formalization of the logic in Isabelle/HOL and automation via symbolically decomposing and  executing HCSP processes according to the logic and the integration with external solvers to handle differential equations and real arithmetic properties. In the future, we will consider to develop more efficient rules for reasoning about ODEs and loops in HHLPar and also apply HHLPar to a wider range of practical case studies.

\section*{Acknowledgements}
This work was supported
 by the National Key R\&D Program of China under grant
 No. 2022YFA1005101, the Natural Science Foundation of
 China (NSFC) under grants No. 62432005 No. 62032024, and No. 62192732.

\bibliographystyle{plain}
\bibliography{arxiv}

\newpage
\appendix
\section{Trace-based HHL}

\subsection{Trace Synchronization}\label{Trace Synchronization}
The full definition of trace synchronization function is defined as following:

{\small
\[
\begin{array}{c}
\prftree[r]{SyncIO}{ch\in cs}{tr_1\|_{cs} tr_2 \Downarrow \textit{tr}}
{\langle ch!,v\rangle^\chop tr_1 \|_{cs} \langle ch?,v\rangle ^\chop tr_2 \Downarrow \langle ch,v\rangle ^\chop \textit{tr}}
\vspace{2mm}\\
\prftree[r]{NoSyncIO}{ch\notin cs}{tr_1\|_{cs} tr_2 \Downarrow \textit{tr}}
{\langle ch\triangleright,v\rangle^\chop tr_1 \|_{cs} tr_2 \Downarrow \langle ch\triangleright,v\rangle^\chop \textit{tr}} 
\quad
\prftree[r]{SyncEmpty1}{ch\in cs}{ }
{\langle ch\triangleright,v\rangle^\chop tr_1 \|_{cs} \epsilon  \Downarrow \delta}
\vspace{2mm}\\
\prftree[r]{SyncEmpty2}{tr_1\|_{cs} \epsilon  \Downarrow \textit{tr}} 
{\langle d,\overrightarrow{p}_1,\rdy_1\rangle^\chop tr_1 \|_{cs} \epsilon  \Downarrow \delta} \quad
\prftree[r]{SyncEmpty3}{\epsilon \|_{\mathit{cs}} \epsilon \Downarrow \epsilon}  
\vspace{2mm}\\
\prftree[r]{SyncWait1 }
{tr_1\|_{cs} tr_2 \Downarrow \textit{tr}}
{\compat(\rdy_1,\rdy_2) \quad d >0 }
{
\begin{array}{c}
     \langle d,\overrightarrow{p}_1,\rdy_1\rangle^\chop tr_1 \|_{cs} \langle d,\overrightarrow{p}_2,\rdy_2\rangle^\chop tr_2 \Downarrow  \\
     \langle d, \overrightarrow{p}_1\uplus \overrightarrow{p}_2,(\rdy_1\cup \rdy_2)-cs\rangle^\chop \textit{tr} 
\end{array}
} 
\vspace{2mm}\\
\prftree[r]{SyncWait2 }
{\begin{array}{c}
		d_1>d_2>0  \quad
		\compat(\rdy_1,\rdy_2)\\
  \langle d_1-d_2,\overrightarrow{p}_1(\cdot+d_2),\rdy_1 \rangle^\chop tr_1\|_{cs} tr_2\Downarrow \textit{tr}
\end{array}}
{\begin{array}{c}
     \langle d_1,\overrightarrow{p}_1,\rdy_1\rangle^\chop tr_1 \|_{cs} \langle d_2,\overrightarrow{p}_2,\rdy_2\rangle^\chop tr_2 \Downarrow  \\
     \langle d_2,\overrightarrow{p}_1\uplus \overrightarrow{p}_2,(\rdy_1\cup \rdy_2)-cs \rangle^\chop \textit{tr} 
\end{array}
}
\end{array}
\]}

\subsection{Big-step Semantics}\label{Big-step Semantics}
The full big-step semantics of HCSP process is defined by the following rules:
{\small
\[
\begin{array}{c}
      \prftree[r]{SkipB}{(\pskip,s) \Rightarrow (s,\epsilon)} 
	\quad 
 \prftree[r]{AssignB}{(x:=e,s) \Rightarrow (s[x \mapsto e],\epsilon)} 
	\vspace{2mm}\\ 
  \prftree[r]{OutB1}{(ch!e, s) \Rightarrow (s, \langle ch!, s(e)\rangle)}
	\quad
 \prftree[r]{OutB2}{(ch!e, s) \Rightarrow (s, \langle d, I_s, \{ch!\}\rangle ^\chop \langle ch!, s(e)\rangle)}  
	\vspace{2mm}\\ 
 \prftree[r]{InB1}{(ch?x, s) \Rightarrow (s[x \mapsto v], \langle ch?, v\rangle) } 
  \quad 
  \prftree[r]{InB2}{(ch?x, s) \Rightarrow (s[x \mapsto v], \langle d, I_s, \{ch?\} \rangle^\chop \langle ch?, v\rangle)} 
	\vspace{2mm}\\
	\prftree[r]{RepB1}{ }
	{(c^*, s) \Rightarrow (s, \epsilon)} 
 \quad
	\prftree[r]{RepB2}{ (c, s)  \Rightarrow (s_1, \textit{tr}_1) \quad (c^*, s_1) \Rightarrow (s_2, \textit{tr}_2)  }
	{ (c^*, s) \Rightarrow (s_2, {\textit{tr}_1}^\chop \textit{tr}_2)} 
 \vspace{2mm}\\
 \prftree[r]{WaitB}{}{(\pwait e,s)\Rightarrow(s,\langle s(e),I_s,\{\})}
	\quad \prftree[r]{SeqB}{(c, s_1) \Rightarrow (s_2, tr_1)}{(c_2, s_2) \Rightarrow (s_3, tr_2)}
	{(c_1; c_2, s_1) \Rightarrow (s_3, {tr_1}^\chop tr_2)}
\end{array}
\]}

{\small
\[
\begin{array}{cc}
   \prftree[r]{CondB1}{s_1(B)}{(c_1, s_1) \Rightarrow (s_2, \textit{tr})}
	{(\IFE{B}{c_1}{c_2}, s_1) \Rightarrow (s_2, \textit{tr})} 
 \quad
  \prftree[r]{IChoiceB1}{(c_1, s_1) \Rightarrow (s_2, \textit{tr})}
	{(c_1 \sqcup c_2, s_1) \Rightarrow (s_2, \textit{tr})} 
	\vspace{2mm}\\
 \prftree[r]{CondB2}{\neg s_1(B)}{(c_2, s_1) \Rightarrow (s_2, \textit{tr})}
	{(\IFE{B}{c_1}{c_2}, s_1) \Rightarrow (s_2, \textit{tr})} 
	\quad
  \prftree[r]{IChoiceB2}{(c_2, s_1) \Rightarrow (s_2, \textit{tr})}
	{(c_1 \sqcup c_2, s_1) \Rightarrow (s_2, \textit{tr})} 
	\vspace{2mm}\\
 \prftree[r]{ContB1}{
		\neg B(s)}
	{(\evo{x}{e}{B}, s) \Rightarrow (s, \epsilon)}
	\vspace{2mm}\\
 \prftree[r]{ContB2}{
		\begin{array}{cc}
			\overrightarrow{p} \mbox{ is a solution of the ODE $\overrightarrow{\dot{x}}=\overrightarrow{e}$} \\
			\overrightarrow{p}(0)=s(\overrightarrow{x})\quad \forall t\in[0,d).\,s[\overrightarrow{x}\mapsto \overrightarrow{p}(t)](B) \quad \neg s[\overrightarrow{x}\mapsto \overrightarrow{p}(d)](B)
	\end{array}}
	{(\evo{x}{e}{B}, s) \Rightarrow (s[\overrightarrow{x}\mapsto \overrightarrow{p}(d)], \langle d, \overrightarrow{p}, \{\}\rangle) }
 \vspace{2mm}\\
\prftree[r]{IntB1}{i\in L}{ch_i*=ch!e}{(c_i,s_1) \Rightarrow (s_2,\textit{tr})}
	{(\exempt{\evo{x}{e}{B\propto c}}{i\in L}{ch_i*}{c_i}, s_1) \Rightarrow
		(s_2, \langle ch!, s_1(e)\rangle^\chop \textit{tr})} 
		\vspace{2mm}\\
  \prftree[r]{IntB2}{
		\begin{array}{cc}
			\overrightarrow{p} \mbox{ is a solution of the ODE $\overrightarrow{\dot{x}}=\overrightarrow{e}$} \quad \overrightarrow{p}(0)=s_1(\overrightarrow{x}) \\
			\forall t\in[0,d).\,s_1[\overrightarrow{x}\mapsto \overrightarrow{p}(t)](B) \\ 
   i\in L\quad ch_i*=ch!e\quad (c_i,s_1[\overrightarrow{x}\mapsto \overrightarrow{p}(d)])\Rightarrow (s_2,\textit{tr})
	\end{array}}
	{\begin{array}{cc}
	     (\exempt{\evo{x}{e}{B\propto c}}{i\in L}{ch_i*}{c_i}, s_1) \Rightarrow  \\
	     (s_2,
		\langle d, \overrightarrow{p}, \rdy(\cup_{i\in L} ch_i*)\rangle^\chop \langle ch!,s_1[\overrightarrow{x}\mapsto \overrightarrow{p}(d)](e)\rangle ^\chop \textit{tr}) 
	\end{array}
  }
  \vspace{2mm}\\
		\prftree[r]{IntB3}{i\in L}{ch_i*=ch?y}{(c_i,s_1[y\mapsto v]) \Rightarrow (s_2,\textit{tr})}
	{(\exempt{\evo{x}{e}{B\propto c}}{i\in L}{ch_i*}{c_i}, s_1) \Rightarrow
		(s_2, \langle ch?, v\rangle^\chop \textit{tr})}
  \vspace{2mm}\\ 
  \prftree[r]{IntB4}{
		\begin{array}{cc}
			\overrightarrow{p} \mbox{ is a solution of the ODE $\overrightarrow{\dot{x}}=\overrightarrow{e}$} \quad \overrightarrow{p}(0)=s_1(\overrightarrow{x}) \\
			\forall t\in[0,d).\,s_1[\overrightarrow{x}\mapsto \overrightarrow{p}(t)](B) \\
   i\in L\quad ch_i*=ch?y\quad (c_i, s_1[\overrightarrow{x}\mapsto \overrightarrow{p}(d),y\mapsto v])\Rightarrow (s_2,\textit{tr})
	\end{array}}
	{\begin{array}{cc}
	      (\exempt{\evo{x}{e}{B\propto c}}{i\in L}{ch_i*}{c_i}, s_1) \Rightarrow \\
	     (s_2,
		\langle d, \overrightarrow{p}, \rdy(\cup_{i\in L} ch_i*)\rangle^\chop \langle ch?,v\rangle ^\chop \textit{tr}) 
	\end{array}
  }\vspace{2mm}\\
  \prftree[r]{IntB5}{\neg s(B)}{(c,s_1)\Rightarrow(s_2,tr)}
	{(\exempt{\evo{x}{e}{B\propto c}}{i\in L}{ch_i*}{c_i}, s1) \Rightarrow (s2, tr)}
\vspace{2mm}\\
	\prftree[r]{IntB6}{
		\begin{array}{cc}
			\overrightarrow{p} \mbox{ is a solution of the ODE $\overrightarrow{\dot{x}}=\overrightarrow{e}$} \quad\overrightarrow{p}(0)=s_1(\overrightarrow{x})\\
			 \forall t\in[0,d).\,s_1[\overrightarrow{x}\mapsto \overrightarrow{p}(t)](B) \quad \neg s_1[\overrightarrow{x}\mapsto \overrightarrow{p}(d)](B) \\
    (c,s_1[\overrightarrow{x}\mapsto \overrightarrow{p}(d)])\Rightarrow(s_2,tr)
	\end{array}}
	{\begin{array}{cc}
	     (\exempt{\evo{x}{e}{B\propto c}}{i\in L}{ch_i*}{c_i}, s_1) \Rightarrow  \\
	     (s_2,\langle d,\overrightarrow{p},\rdy(\cup_{i\in L} ch_i*)\rangle^\chop tr) 
	\end{array}}
	\vspace{2mm}\\
	\prftree[r]{ParB}{(c_1, s_1) \Rightarrow (s_1', tr_1)}{(c_2, s_2) \Rightarrow (s_2', tr_2)}{tr_1\|_{cs}tr_2 \Downarrow \textit{tr}}
{(c_1 \|_{cs} c_2, s_1\uplus s_2) \Rightarrow (s_1'\uplus s_2', \textit{tr})}  
\end{array}
\]}

\section{Complement Sequential Rules}\label{App:seq}

\subsection{}
In this section, we first explain the rules for interrupt command with explicit solution in detail.

Given an interrupt command $\langle \overrightarrow{\dot{x}}=\overrightarrow{e}\& B\propto c'\rangle \unrhd \talloblong_{i\in L} (ch_i* \rightarrow c_i)$, where we use $es$ to denote the list of communications in the form $(ch?x\rightarrow c_i)$ or $(ch!e\rightarrow c_i)$, and $f$ is a solution to $\overrightarrow{\dot{x}}=\overrightarrow{e}$, the branches of assertions corresponding to the communication list  is computed by $\mathsf{rel\_cm}(es,c,f)$, if for each $es[i]=(ch?y\rightarrow c_i)$, we have $\mathsf{spec\_of}(c_i;c,Q_i)$ then 
{\small
\[
\mathsf{rel\_cm}(es,c,f)[i]=\langle ch?, \{\mathsf{d,v}\Rightarrow Q_i[y := \mathsf{v}][\overrightarrow{x}:=f(s_0(\overrightarrow{x}),\mathsf{d})]\} \rangle 
\]}
\par\noindent
and for each $es[i]=(ch!e\rightarrow c_i)$, we have $\mathsf{spec\_of}(c_i;c,Q_i)$ then
{\small
\[
\mathsf{rel\_cm}(es,c,f)[i]=\langle ch!,\{\mathsf{d}\Rightarrow e(p(s_0,\mathsf{d}))\} ,\{\mathsf{d}\Rightarrow Q_i[\overrightarrow{x}:=f(s_0(\overrightarrow{x}),\mathsf{d})] \} \rangle 
\]}
\par\noindent
Then the inference rule for interrupt is:
{\small
\[
\prftree
{\begin{array}{c}
     \mathsf{paramODEsol}(\overrightarrow{\dot{x}}=\overrightarrow{e},B,f,e)  \qquad
     \mathsf{lipschitz}(\overrightarrow{\dot{x}}=\overrightarrow{e})
     \\
     \mathsf{spec\_of}(c';c,P) \qquad
     \forall\,i\in L,\,\mathsf{spec\_of}(c_i;c,Q_i)
\end{array}}
{
\begin{array}{cc}
     &  \mathsf{spec\_of}(\langle \overrightarrow{\dot{x}}=\overrightarrow{e}\& B\propto c'\rangle \unrhd \talloblong_{i\in L} (ch_i* \rightarrow c_i);c, \mathsf{interrupt}(\overrightarrow{x}\rightarrowtail f(\overrightarrow{x},t),\\
     & e,\{\mathsf{d}\Rightarrow P[\overrightarrow{x}:=f(s_0(\overrightarrow{x}),\mathsf{d})]\},\mathsf{rel\_cm}(es,c,f))
\end{array}
}
\]}
\par\noindent
The meaning of this rule is as follows: the specification of the interrupt first evolves along the path $p(t)=s_0[\overrightarrow{x}\mapsto f(s_0(\overrightarrow{x}),t)]$, and one of the following three situations occurs:
\begin{itemize}
    \item If the evolution is interrupted by an input communication $(ch?x\rightarrow c_i)$ at time $d$ and with value $v$, then update the state to $s_0[\overrightarrow{x}\mapsto f(s_0(\overrightarrow{x}),d)][x \mapsto v]$, followed by the behavior of $c_i;c$ as specified by $Q_i$.
    
    \item If the evolution is interrupted by an output communication $(ch!e\rightarrow c_i)$ at time $d$ and with value $v=e(s_0[\overrightarrow{x}\mapsto f(s_0(\overrightarrow{x}),d)])$, and then update the  state to $s_0[\overrightarrow{x}\mapsto f(s_0(\overrightarrow{x}),d)]$, followed by the behavior of $c_i;c$ as specified by $Q_i$.

    \item If no interrupt occurs before time $d=s_0(e)$, then update the state to $s_0[\overrightarrow{x}\mapsto f(s_0(\overrightarrow{x}),d)]$, followed by the behavior of $c';c$ as specified by $P$.
\end{itemize}

The above assumes that the ODE with boundary condition has a solution of finite length for any starting state. Another important case is when the ODE has a solution of infinite length, in particular when the boundary condition is $\mathsf{true}$. In this case, the appropriate assertion is $\mathsf{interrupt_\infty}$. We first define predicate $\mathsf{paramODEsolInf}(\overrightarrow{\dot{x}}=\overrightarrow{e},f)$, meaning that $f$ is the (infinite length) solution to $\overrightarrow{\dot{x}}=\overrightarrow{e}$, then the corresponding rule is:
{\small
\[
\prftree{\mathsf{paramODEsolInf}(\overrightarrow{\dot{x}}=\overrightarrow{e},\overrightarrow{p})}{\mathsf{lipschitz}(\overrightarrow{\dot{x}}=\overrightarrow{e})}
{\forall\,i\in L,\,\mathsf{spec\_of}(c_i;c,Q_i)}
{\begin{array}{cc}
    & \mathsf{spec\_of}(\langle\overrightarrow{\dot{x}}=\overrightarrow{e}\& \mathsf{true}\propto c'\rangle \unrhd \talloblong_{i\in L} (ch_i* \rightarrow c_i);c,\\
    & \mathsf{interrupt_\infty}(\overrightarrow{x}\rightarrowtail f(\overrightarrow{x},t), \mathsf{rel\_cm}(es,c,f))
\end{array}
}
\]}

Next, we introduce the rules for interrupt with differential invariants.

Similarly, we define the branches of assertions corresponding to the communication list, denoted by $\mathsf{relinv\_cm}(es,c,inv)$, if for each $es[i]=(ch?y\rightarrow c_i)$, we have $\mathsf{spec\_of}(c_i;c,Q_i)$ then 
{\small
\[
\mathsf{relinv\_cm}(es,c,inv)[i]=\langle ch?, \{\mathsf{d,v}\Rightarrow (\uparrow inv\wedge Q_i[y := \mathsf{v}])[\overrightarrow{x}:=\overrightarrow{nx_i}]\} \rangle
\]}
\par\noindent
and for each $es[i]=(ch!e\rightarrow c_i)$, we have $\mathsf{spec\_of}(c_i;c,Q_i)$ then
{\small
\[
\mathsf{relinv\_cm}(es,c,inv)[i]=\langle ch!,\{\mathsf{d}\Rightarrow e(s_0[\overrightarrow{x}:=\overrightarrow{nx_i}])\} ,\{\mathsf{d}\Rightarrow (\uparrow inv\wedge Q_i)[\overrightarrow{x}:=\overrightarrow{nx_i}] \} \rangle 
\]}
\par\noindent
And then, we have the following rule:
{\small
\[
\prftree{
\begin{array}{c}
\mathsf{paramODEInv}(\overrightarrow{\dot{x}}=\overrightarrow{e}, B, inv, pp)\quad\mathsf{lipschitz}(\overrightarrow{\dot{x}}=\overrightarrow{e}) \\
    \mathsf{spec\_of}(c';c,P)\quad  \forall\,i\in L,\,\mathsf{spec\_of}(c_i;c,Q_i)
\end{array}
}
{\begin{array}{c}
     \mathsf{spec\_of}(\langle \overrightarrow{\dot{x}}=\overrightarrow{e}\& B\propto c'\rangle \unrhd \talloblong_{i\in L} (ch_i* \rightarrow c_i);c, (\uparrow(\neg B)\bar{\wedge} P)\bar{\vee}\uparrow(\neg pp\wedge B)\bar{\vee}\\
     \exists\,T\,\overrightarrow{nx}\,\overrightarrow{nx_i}_{\in L}.\,(\uparrow (pp\wedge B)\bar{\wedge}\mathsf{interrupt}(
    inv, T, \\
    \{d \Rightarrow (\uparrow(inv\wedge bound(B)) \bar{\wedge} P)[\overrightarrow{x}:=\overrightarrow{nx}] \},\mathsf{relinv\_cm}(es,c,inv))) )
\end{array}}
\]}

If the ODE in interrupt command has infinite length, we have:
{\small
\[
\prftree{\mathsf{paramODEInv}(\overrightarrow{\dot{x}}=\overrightarrow{e}, B, inv, pp)}{\mathsf{lipschitz}(\overrightarrow{\dot{x}}=\overrightarrow{e})}{\forall\,i\in L,\,\mathsf{spec\_of}(c_i;c,Q_i)}
{\begin{array}{c}
     \mathsf{spec\_of}(\langle \overrightarrow{\dot{x}}=\overrightarrow{e}\& \mathsf{true}\propto c'\rangle \unrhd \talloblong_{i\in L} (ch_i* \rightarrow c_i);c, \uparrow(\neg pp)\bar{\vee}\\
     \exists\,\overrightarrow{nx_i}_{\in L}.\,(\uparrow pp\bar{\wedge}\mathsf{interrupt_\infty}(
    inv, \mathsf{relinv\_cm}(es,c,inv))) )
\end{array}}
\]}

\subsection{}
In this section we give the all the sequential rules without subsequent process.

{\small
\[\begin{array}{c}
   \prftree
{\mathsf{spec\_of}(x := e, \mathsf{init}[x := e])}   \vspace{2mm} \\
\prftree{\mathsf{spec\_of}(c_1, P)}{\mathsf{spec\_of}(c_2, Q)}
{\mathsf{spec\_of}(\IFE{B}{c_1}{c_2}, (\uparrow(B)\bar{\land} P) \bar{\lor} (\uparrow(\neg B)\bar{\land} Q))}\vspace{2mm} \\
\prftree{\mathsf{spec\_of}(ch?x, \mathsf{wait\_in}(\mathsf{id\_inv}, ch, \{\mathsf{d,v}\Rightarrow \mathsf{init}[x := \mathsf{v}]\}))}\vspace{2mm} \\
\prftree{\mathsf{spec\_of}(ch!e, \mathsf{wait\_outv}(\mathsf{id\_inv}, ch, e, \{\mathsf{d}\Rightarrow \mathsf{init}\}))} \vspace{2mm} \\
\prftree{\mathsf{spec\_of}(\pwait\ e, \mathsf{wait}(\mathsf{id}, e, \{\mathsf{d}\Rightarrow \mathsf{init}\}))} \vspace{2mm} \\ 
\prftree{\mathsf{paramODEsol}(\overrightarrow{\dot{x}}=\overrightarrow{e}, B, f, e)}{\mathsf{lipschitz}(\overrightarrow{\dot{x}}=\overrightarrow{e})}
{\mathsf{spec\_of}(\langle \overrightarrow{\dot{x}}=\overrightarrow{e}\& B\rangle, \mathsf{wait}(
\overrightarrow{x}\rightarrowtail f(\overrightarrow{x},t), e, \{\mathsf{d} \Rightarrow \mathsf{init}[\overrightarrow{x}:=f(s_0(\overrightarrow{x}),\mathsf{d})] \}))} \vspace{2mm} \\    
\prftree{\forall\ d\ Q.\ \mathsf{spec\_of}(d,Q) \longrightarrow \mathsf{spec\_of}(c;d,F(Q))}
{\mathsf{spec\_of}(c^*,\mathsf{Rec}\ R.\  \mathsf{init}\bar{\lor} F(R))}\vspace{2mm}\\
\prftree
{\begin{array}{c}
     \mathsf{paramODEsol}(\overrightarrow{\dot{x}}=\overrightarrow{e},B,f,e) \qquad
     \mathsf{lipschitz}(\overrightarrow{\dot{x}}=\overrightarrow{e})\\
     \mathsf{spec\_of}(c',P)\qquad \forall\,i\in L.\, \mathsf{spec\_of}(c_i,Q_i)
\end{array}}
{
\begin{array}{cc}
     &  \mathsf{spec\_of}(\langle \overrightarrow{\dot{x}}=\overrightarrow{e}\& B\propto c'\rangle \unrhd \talloblong_{i\in L} (ch_i* \rightarrow c_i), \mathsf{interrupt}(\overrightarrow{x}\rightarrowtail f(\overrightarrow{x},t),\\
     & e,\{\mathsf{d}\Rightarrow \mathsf{P}[\overrightarrow{x}:=f(s_0(\overrightarrow{x}),\mathsf{d})]\},\mathsf{rel\_cm}(es,\pskip,f))
\end{array}
}
\vspace{2mm}\\
\prftree{\mathsf{paramODEsolInf}(\overrightarrow{\dot{x}}=\overrightarrow{e},f)}{\mathsf{lipschitz}(\overrightarrow{\dot{x}}=\overrightarrow{e})}
{
\begin{array}{cc}
     &  \mathsf{spec\_of}(\langle \overrightarrow{\dot{x}}=\overrightarrow{e}\& \mathit{true}\propto c'\rangle\unrhd \talloblong_{i\in L} (ch_i* \rightarrow c_i), \mathsf{interrupt_\infty}(\overrightarrow{x}\rightarrowtail f(\overrightarrow{x},t), \\
     & \mathsf{rel\_cm}(es,\pskip,f))
\end{array}
}
\vspace{2mm}\\
\prftree{\mathsf{paramODEInv}(\overrightarrow{\dot{x}}=\overrightarrow{e}, B, inv, pp)}{\mathsf{lipschitz}(\overrightarrow{\dot{x}}=\overrightarrow{e})}
{
\begin{array}{c}
     \mathsf{spec\_of}(\langle \overrightarrow{\dot{x}}=\overrightarrow{e}\& B\rangle;c, (\uparrow(\neg B)\bar{\wedge} \mathsf{init})\bar{\vee}\uparrow(\neg pp\wedge B)\bar{\vee}\\
     \exists\,T\,\overrightarrow{nx}.\,(\uparrow (pp\wedge B)\bar{\wedge}\mathsf{wait}(
    inv, T, \{\mathsf{d} \Rightarrow (\uparrow(inv\wedge bound(B)) \bar{\wedge} \mathsf{init})[\overrightarrow{x}:=\overrightarrow{nx}] \})) )
\end{array}
}
\vspace{2mm}\\
\prftree{
\begin{array}{c}
     \mathsf{paramODEInv}(\overrightarrow{\dot{x}}=\overrightarrow{e}, B, inv, pp) \quad \mathsf{lipschitz}(\overrightarrow{\dot{x}}=\overrightarrow{e})\\
     \mathsf{spec\_of}(c',P)\quad \forall\,i\in L,\,\mathsf{spec\_of}(c_i,Q_i)
\end{array}
}
{\begin{array}{c}
     \mathsf{spec\_of}(\langle \overrightarrow{\dot{x}}=\overrightarrow{e}\& B\propto c'\rangle \unrhd \talloblong_{i\in L} (ch_i* \rightarrow c_i), (\uparrow(\neg B)\bar{\wedge} P)\bar{\vee}\uparrow(\neg pp\wedge B)\bar{\vee}\\
     \exists\,T\,\overrightarrow{nx}\,\overrightarrow{nx_i}_{\in L}.\,(\uparrow (pp\wedge B)\bar{\wedge}\mathsf{interrupt}(
    inv, T, \\
    \{\mathsf{d} \Rightarrow (\uparrow(inv\wedge bound(B)) \bar{\wedge} P)[\overrightarrow{x}:=\overrightarrow{nx}] \},\mathsf{relinv\_cm}(es,\pskip,inv))) )
\end{array}}
\vspace{2mm}\\
\prftree{\mathsf{paramODEInv}(\overrightarrow{\dot{x}}=\overrightarrow{e}, B, inv, pp)}{\mathsf{lipschitz}(\overrightarrow{\dot{x}}=\overrightarrow{e})}{\forall\,i\in L,\,\mathsf{spec\_of}(c_i,Q_i)}
{\begin{array}{c}
     \mathsf{spec\_of}(\langle \overrightarrow{\dot{x}}=\overrightarrow{e}\& \mathsf{true}\propto c'\rangle \unrhd \talloblong_{i\in L} (ch_i* \rightarrow c_i), \uparrow(\neg pp)\bar{\vee}\\
     \exists\,\overrightarrow{nx_i}_{\in L}.\,(\uparrow pp\bar{\wedge}\mathsf{interrupt_\infty}(
    inv, \mathsf{relinv\_cm}(es,c,inv))) )
\end{array}}
\end{array}
\]}

\section{Complement Synchronization Rules}
In this section we show the other synchronization rules.
\label{App:par}

First, we introduce the rules involving the common operators of assertions. 
{\small
\[
\prftree[r]{False}{\mathsf{sync}(chs,\mathsf{false},P)(s_0)\Longrightarrow_a \mathsf{false}(s_0)}
\]}
\par\noindent
if one side is a $\mathsf{false}$ assertion, we obtain a result of $\mathsf{false}$.
{\small
\[
\prftree[r]{Disj}{\mathsf{sync}(chs,P_1,Q)(s_0)\Longrightarrow_a R_1(s_0)}
{\mathsf{sync}(chs,P_2,Q)(s_0)\Longrightarrow_a R_2(s_0)}
{\mathsf{sync}(chs,P_1\bar{\lor} P_2,Q)(s_0)\Longrightarrow_a (R_1\bar{\lor} R_2)(s_0)}
\]}
\par\noindent
if one side is a disjunction, we can eliminate this to its components.
{\small
\[
\prftree[r]{Bool}
{b(s_1)\longrightarrow\mathsf{sync}(chs,P,Q)(s_0)\Longrightarrow_a R(s_0)}
{\mathsf{sync}(chs,\uparrow b\bar{\land} P,Q)(s_0)\Longrightarrow_a (\uparrow b \bar{\land} R)(s_0)}
\]}
\par\noindent
if one side is a conjunction with a boolean expression $b$, we perform synchronization on the rest part under $b$ and pull out $b$ lifted on a parallel state as a new condition. 
{\small
\[
\prftree[r]{Subst}
{\mathsf{sync}(chs,P[x:=e],Q)(s_0)\Longrightarrow_a \mathsf{sync}(chs,P,Q)[x:=e](s_0)}
\]} 
\par\noindent
if one side is a substitution assertion, the substitution can be pulled out after lifting.

In principle, $\mathsf{wait\_out}$, $\mathsf{wait\_in}$ and $\mathsf{wait}$ are all special cases of $\mathsf{interrupt}$ (including $\mathsf{interrupt_\infty}$, by viewing $\mathsf{interrupt_\infty}$(I,cm) as $\mathsf{interrupt}(I,\infty,\{\mathsf{d}\Rightarrow \mathsf{false}\},cm)$).
Thus, the synchronization rule for interrupt assertion is complex and contains all the potential situations. We will first give some simple cases, and then introduce the rule for $\mathsf{interrupt}$ as a complete form.

While synchronizing two init assertions, we can easily infer that the state of each part remains the same and the traces on both sides are empty lists. Naturally, we have
{\small
\[
\prftree[r]{InitInit}{\mathsf{sync}(chs,\mathsf{init},\mathsf{init})(s_0)\Longrightarrow_a \mathsf{init}(s_0)}
\]
}

While synchronizing an init assertion and an wait assertion, if the wait time is greater than 0, we directly obtain a false assertion. Otherwise, if the wait time is Less than or equal to 0, the wait assertion turns to its tail by the definition.
{\small
\[
\prftree[r]{WaitInit}{\mathsf{sync}(chs,\mathsf{wait}(I,e,\{\mathsf{d}\Rightarrow P\}),\mathsf{init})(s_0)\Longrightarrow_a \uparrow(e\le 0)\bar{\land}\mathsf{sync}(chs,P|_{\mathsf{d}=0},\mathsf{init})(s_0)}
\]
}

While synchronizing an init assertion and an input assertion, if the communication channel belongs to the common channel set, we directly obtain a false assertion, Otherwise, this external communication must occur at once, since the init assertion does not support any waiting time. Thus, we have:
{\small
\[
\prftree[r]{InInit1}{ch\in chs}{\mathsf{sync}(chs,\mathsf{wait\_in}(I,ch,\{\mathsf{d}\Rightarrow P\}),\mathsf{init})(s_0)\Longrightarrow_a \mathsf{false}(s_0)}
\]
\[
\prftree[r]{InInit2}{ch\notin chs}
{
\begin{array}{c}
     \mathsf{sync}(chs,\mathsf{wait\_in}(I,ch,\{\mathsf{d,v}\Rightarrow P\}),\mathsf{init})(s_0)\Longrightarrow_a 
     \mathsf{interrupt}(I\uplus\mathsf{id},0,\\
     \{\mathsf{d}\Rightarrow\mathsf{false}\},[\langle ch?,\mathsf{d,v}\Rightarrow\mathsf{sync}(chs,P,\mathsf{init})\rangle])(s_0)
\end{array}
}
\]
}
\par\noindent
The rules for synchronizing an init assertion and an output assertion are similar.

While synchronizing an output assertion and an input assertion, we need to consider the different cases of whether their channels belong to the common channel set. If they are both in the set and have the same name, then the handshake occurs at once. while if they have different names which means both sides are waiting for a handshake, but they don't match and this lead to a deadlock represented by a false assertion. So we have the following rules:
{\small
\[
\prftree[r]{InOut1}{ch_1\in chs}{ch_2\in chs}{ch_1=ch_2}{
\begin{array}{c}
     \mathsf{sync}(chs,\mathsf{wait\_in}(I_1,ch_1,\{\mathsf{d_1,v_1}\Rightarrow P_1\}),\mathsf{wait\_outv}(I_2,ch_2,e,\{\mathsf{d_2}\Rightarrow P_2\}))(s_0) \\\Longrightarrow_a 
     \mathsf{sync}(chs,P_1|_{\mathsf{d_1}=0,\mathsf{v_1}=s_0(e)},P_2|_{\mathsf{d_2}=0})(s_0)
\end{array}
}
\]
\[
\prftree[r]{InOut2}{ch_1\in chs}{ch_2\in chs}{ch_1\neq ch_2}
{
\begin{array}{c}
     \mathsf{sync}(chs,\mathsf{wait\_in}(I_1,ch_1,\{\mathsf{d_1,v_1}\Rightarrow P_1\}),\mathsf{wait\_outv}(I_2,ch_2,e,\{\mathsf{d_2}\Rightarrow P_2\}))(s_0)\\\Longrightarrow_a  
     \mathsf{false}(s_0) 
\end{array}
}
\]
}
\par\noindent
If at least one of them is an external communication, then it must happen before the internal communication, because the condition for the internal handshake to occur are not met. Thus, we have:
{\small
\[
\prftree[r]{InOut3}{ch_1\in chs}{ch_2\notin chs}{
\begin{array}{c}
     \mathsf{sync}(chs,\mathsf{wait\_in}(I_1,ch_1,\{\mathsf{d_1,v_1}\Rightarrow P_1\}),\mathsf{wait\_outv}(I_2,ch_2,e,\{\mathsf{d_2}\Rightarrow P_2\}))(s_0)\\\Longrightarrow_a  
     \mathsf{wait\_outv}(I_1\uplus I_2,ch_2,e,\{\mathsf{d_2}\Rightarrow\\
     \mathsf{sync}(chs,\mathsf{wait\_in}(I_1|_{t=t+\mathsf{d_2}},ch_1,\{\mathsf{d_1,v_1}\Rightarrow P_1|_{\mathsf{d_1=d_1+d_2}}\}),P_2))\}
     )(s_0)
\end{array}
}
\]
\[
\prftree[r]{InOut4}{ch_1\notin chs}{ch_2\in chs}{
\begin{array}{c}
     \mathsf{sync}(chs,\mathsf{wait\_in}(I_1,ch_1,\{\mathsf{d_1,v_1}\Rightarrow P_1\}),\mathsf{wait\_outv}(I_2,ch_2,e\{\mathsf{d_2}\Rightarrow P_2\}))(s_0)\\\Longrightarrow_a  
     \mathsf{wait\_in}(I_1\uplus I_2,ch_1,\{\mathsf{d_1,v_1}\Rightarrow\\
     \mathsf{sync}(chs,P_1,\mathsf{wait\_outv}(I_2|_{t=t+\mathsf{d_1}},ch_2,e,\{\mathsf{d_2}\Rightarrow P_2|_{\mathsf{d_2=d_2+d_1}}\}))\}
     )(s_0)
\end{array}
}
\]
\[
\prftree[r]{InOut5}{ch_1\notin chs}{ch_2\notin chs}{
\begin{array}{c}
     \mathsf{sync}(chs,\mathsf{wait\_in}(I_1,ch_1,\{\mathsf{d_1,v_1}\Rightarrow P_1\}),\mathsf{wait\_outv}(I_2,ch_2,e,\{\mathsf{d_2}\Rightarrow P_2\}))(s_0)\\\Longrightarrow_a  
     \mathsf{interrupt_\infty}(I_1\uplus I_2,
     [\langle ch_1?,\{\mathsf{d_1,v_1}\Rightarrow
     \mathsf{sync}(chs,\\
     P_1,\mathsf{wait\_outv}(I_2|_{:=t+\mathsf{d_1}},ch_2,e,\{\mathsf{d_2}\Rightarrow P_2|_{\mathsf{d_1=d_1+d_2}}\}))\}\rangle,\\
     \langle ch2!,\{\mathsf{d_2}\Rightarrow e\},
     \{\mathsf{d_2}\Rightarrow
     \mathsf{sync}(chs,\\
     \mathsf{wait\_in}(I_1|_{t=t+\mathsf{d_2}},ch_1,\{\mathsf{d_1,v_1}\Rightarrow P_1|_{\mathsf{d_1=d_1+d_2}}\}),P_2)\}\rangle]
     )(s_0)
\end{array}
}
\]
}

Next, we consider synchronizing two interrupt assertions $\mathsf{interrupt}(I_1,e_1,\{\mathsf{d_1}\Rightarrow P_1\},cm_1)$ and $\mathsf{interrupt}(I_2,e_2,\{\mathsf{d_2}\Rightarrow P_2\},cm_2)$. First, we need to determine whether there is a communication between two sides. The method of judgement is to check if there exists a channel name in the set $chs$, where its input is in the $rdy$ set on one side and its output is in the $rdy$ set on the other side. Define predicate $\mathsf{compat}$ to be the negation of this condition:
{\small
\[
\begin{array}{ll}
     \mathsf{compat}(rdy(cm_1),rdy(cm_2))\triangleq \neg & 
(\exists ch\in chs.(ch!\in rdy(cm_1) \land ch?\in rdy(cm_2)) \\
     & \qquad\lor (ch?\in rdy(cm_1) \land ch!\in rdy(cm_2)))
\end{array}
\]}
\par\noindent
In the case where this predicate holds true, both sides are waiting to be interrupted by external communication, thus its synchronization result should still be in the form of interrupt assertion, and its maximum waiting time is the smaller of $e_1$ and $e_2$. While reaching the maximum waiting time, the shorter one will behave as the tail part and the longer one stays in an incomplete interrupt assertion denoted as $\mathsf{delay}(h,\mathsf{interrupt}(I,e,\{\mathsf{d}\Rightarrow P\},cm))$:
{\small
\[
\begin{array}{ll}
       \mathsf{delay}(h,\mathsf{interrupt}(I,e,\{\mathsf{d}\Rightarrow P\},cm)) \triangleq 
       &
       \mathsf{interrupt}(I|_{t=t+h},e-h,\\
     & \qquad \{\mathsf{d}\Rightarrow P|_{\mathsf{d=d}+h}\},\mathsf{delay\_cm}(cm,h))
\end{array}
\]}
\par\noindent
where for input $cm[i] = \langle ch?,\{\mathsf{d,v}\Rightarrow Q_1\}\rangle$ or output  $cm[i] = \langle ch!,g,\{\mathsf{d}\Rightarrow Q_2\}\rangle$, we have:
{\small
\[
\begin{array}{ll}
     & \mathsf{delay\_cm}(cm,h)[i] = \langle ch?,\{\mathsf{d,v}\Rightarrow Q_1|_{\mathsf{d=d}+h}\}\rangle \\
     & \mathsf{delay\_cm}(cm,h)[i] = \langle ch!,\{\mathsf{d}\Rightarrow g(\mathsf{d}+h)\},\{\mathsf{d}\Rightarrow Q_2|_{\mathsf{d=d}+h}\}\rangle
\end{array}
\]}
\par\noindent
we can easily find that $\mathsf{delay}(0,\mathsf{interrupt}(I,e,\{\mathsf{d}\Rightarrow P\},cm)) = \mathsf{interrupt}(I,e,\{\mathsf{d}\Rightarrow P\},cm)$.
By performing synchronization on them, we get the new tail assertion. 
A potential external interruption from $cm_1$ or $cm_2$ that does not belong to the shared set $chs$ may occur during the waiting. Then, one side will behave as the corresponding assertion recorded in $cm_1$ or $cm_2$, the other side will remain its incomplete interrupt assertion. For this case, the synchronization produces the new communication list composed of two parts: 
$\mathsf{rel1}(cm_1|_{chs^{c}},\mathsf{interrupt}(I_2,e_2,\{\mathsf{d_2}\Rightarrow P_2\},cm_2))$ and $\mathsf{rel2}(cm_2|_{chs^{c}},\mathsf{interrupt}(I_1,e_1,\{\mathsf{d_1}\Rightarrow P_1\},cm_1))$
where $cm_1|_{chs^{c}}$ and $cm_2|_{chs^{c}}$ are lists of communications not in $chs$ extracted from $cm_1$ and $cm_2$. The list functions $\mathsf{rel1}$ and $\mathsf{rel2}$ are set as: if $cm[i]=\langle ch?,\{\mathsf{d,v}\Rightarrow Q_1\}\rangle$,
{\small
\[
\begin{array}{cc}
     &  \mathsf{rel1}(cm,R)[i] = \langle ch?,\{\mathsf{d,v}\Rightarrow\mathsf{sync}(chs,Q_1,\mathsf{delay}(\mathsf{d},R))\}\\
     &  \mathsf{rel2}(cm,R)[i] = \langle ch?,\{\mathsf{d,v}\Rightarrow\mathsf{sync}(chs,\mathsf{delay}(\mathsf{d},R),Q_1)\}
\end{array}
\]}
\par\noindent
if $cm[i]=\langle ch!,g,\{\mathsf{d}\Rightarrow Q_2\}\rangle$,
{\small
\[
\begin{array}{cc}
     &  \mathsf{rel1}(cm,R)[i] = \langle ch!,
     \{\mathsf{d}\Rightarrow g(\mathsf{d})\},
     \{\mathsf{d}\Rightarrow\mathsf{sync}(chs,Q_2,\mathsf{delay}(\mathsf{d},R))\}\\
     &  \mathsf{rel2}(cm,R)[i] = \langle ch!,
     \{\mathsf{d}\Rightarrow g(\mathsf{d})\},
     \{\mathsf{d}\Rightarrow\mathsf{sync}(chs,\mathsf{delay}(\mathsf{d},R),Q_2)\}
\end{array}
\]}

So far we can obtain the following rules:
{\small
\[
\begin{array}{c}
     \prftree[r]{IntInt1}{e1(s_1)<e2(s_2)\land e_2(s_2)>0}{\mathsf{compat}(rdy(cm_1),rdy(cm_2))}
{
\begin{array}{cc}
     &  \mathsf{sync}(chs,\mathsf{interrupt}(I_1,e_1,\{\mathsf{d_1}\Rightarrow P_1\},cm_1),\mathsf{interrupt}(I_2,e_2,\{\mathsf{d_2}\Rightarrow P_2\},cm_2))\\
     & (s_1\uplus s_2)\Longrightarrow_a
     \mathsf{interrupt}(I_1\uplus I_2,e_1,\{\mathsf{d_1}\Rightarrow \\
     &\mathsf{sync}(chs,P_1,\mathsf{delay}(\mathsf{d_1},\mathsf{interrupt}(I_2,e_2,\{\mathsf{d_2}\Rightarrow P_2\},cm_2)))\},\\
     & \mathsf{rel1}(cm_1|_{chs^{c}},\mathsf{interrupt}(I_2,e_2,\{\mathsf{d_2}\Rightarrow P_2\},cm_2)) @\\
     & \mathsf{rel2}(cm_2|_{chs^{c}},\mathsf{interrupt}(I_1,e_1,\{\mathsf{d_1}\Rightarrow P_1\},cm_1)))(s_1\uplus s_2)
\end{array}
}  \vspace{2mm}\\
\prftree[r]{IntInt2}{e1(s_1)=e2(s_2)\lor (e_1(s_1)\le0\land e_2(s_2)\le0)}{\mathsf{compat}(rdy(cm_1),rdy(cm_2))}
{
\begin{array}{cc}
     &  \mathsf{sync}(chs,\mathsf{interrupt}(I_1,e_1,\{\mathsf{d_1}\Rightarrow P_1\},cm_1),\mathsf{interrupt}(I_2,e_2,\{\mathsf{d_2}\Rightarrow P_2\},cm_2))\\
     & (s_1\uplus s_2)\Longrightarrow_a
     \mathsf{interrupt}(I_1\uplus I_2,e_1,\{\mathsf{d}\Rightarrow \\
     & \mathsf{sync}(chs,P_1|_{\mathsf{d_1=d}},\mathsf{delay}(\mathsf{d},\mathsf{interrupt}(I_2,e_2,\{\mathsf{d_2}\Rightarrow P_2\},cm_2)))\\
     &\bar{\lor}\ \mathsf{sync}(chs,\mathsf{delay}(\mathsf{d},\mathsf{interrupt}(I_1,e_1,\{\mathsf{d_1}\Rightarrow P_1\},cm_1)),P_2|_{\mathsf{d_2=d}})\},\\
     & \mathsf{rel1}(cm_1|_{chs^{c}},\mathsf{interrupt}(I_2,e_2,\{\mathsf{d_2}\Rightarrow P_2\},cm_2)) @ \\
     &\mathsf{rel2}(cm_2|_{chs^{c}},\mathsf{interrupt}(I_1,e_1,\{\mathsf{d_1}\Rightarrow P_1\},cm_1)))(s_1\uplus s_2)
\end{array}
}
\end{array}
\]}
\par\noindent
Note that in the definition of interrupt assertion, if the expression of waiting time calculated as a negative value then it has equivalent meaning with 0. That is why we need to compare the expression with 0.

In the case when the $\mathsf{compat}$ function is false, there are three possible scenarios. The first is nondeterministicly executing one of the possible handshakes among all that could occur which we represent as $\mathsf{comm}(cm_1,cm_2)$. It is a disjunction of $\mathsf{sync}(chs,Q_1|_{\mathsf{d_1}=0,\mathsf{v_1}=g(0)},Q_2|_{\mathsf{d_2}=0})$ and  $\mathsf{sync}(chs,Q_1|_{\mathsf{d_1}=0},Q_2|_{\mathsf{d_2}=0,\mathsf{v_2}=g(0)}))$ for
all the pairs satisfying one of the following conditions:
{\small
\[
\begin{array}{c}
     ch\in chs \land  cm_1[i]=\langle ch?,\{\mathsf{d_1,v_1}\Rightarrow Q_1\}\rangle  \land cm_2[j]=\langle ch!,g,\{\mathsf{d_2}\Rightarrow Q_2\}\rangle  \vspace{1mm}\\
     ch\in chs \land  cm_1[i]=\langle ch!,g,\{\mathsf{d_1}\Rightarrow Q_1\}\rangle \land cm_2[j]=\langle ch?,\{\mathsf{d_2,v_2}\Rightarrow Q_2\}\rangle 
\end{array}
\]}
\par\noindent
The second is that if the maximum waiting time $e_1$ or $e_2$ is less than 0, then the corresponding side may immediately transit to the tail assertion. The last one is there is an external interrupt occurring at time 0. We obtain the following rule:
{\small
\[
\prftree[r]{IntInt3}{\neg\mathsf{compat}(rdy(cm_1),rdy(cm_2))}
{
\begin{array}{cc}
     &  \mathsf{sync}(chs,\mathsf{interrupt}(I_1,e_1,\{\mathsf{d_1}\Rightarrow P_1\},cm_1),\mathsf{interrupt}(I_2,e_2,\{\mathsf{d_2}\Rightarrow P_2\},cm_2))\\& (s_1\uplus s_2)\Longrightarrow_a
     \mathsf{interrupt}(I_1\uplus I_2,0,\{\mathsf{d}\Rightarrow \mathsf{comm}(cm_1,cm_2)\bar{\lor}\\
     &(\uparrow (e_1\le0)\bar{\land}\mathsf{sync}(chs,P_1|_{\mathsf{d_1}=0},\mathsf{interrupt}(I_2,e_2,\{\mathsf{d_2}\Rightarrow P_2\},cm_2)))\bar{\lor}\\
     &(\uparrow (e_2\le0)\bar{\land}\mathsf{sync}(chs,\mathsf{interrupt}(I_1,e_1,\{\mathsf{d_1}\Rightarrow P_1\},cm_1),P_2|_{\mathsf{d_2}=0}))\},\\
     & \mathsf{rel1}(cm_1|_{chs^{c}},\mathsf{interrupt}(I_2,e_2,\{\mathsf{d_2}\Rightarrow P_2\},cm_2)) @ \\
     &\mathsf{rel2}(cm_2|_{chs^{c}},\mathsf{interrupt}(I_1,e_1,\{\mathsf{d_1}\Rightarrow P_1\},cm_1)))(s_1\uplus s_2)
\end{array}
}
\]}

.

While synchronizing an interrupt assertion and an init assertion (representing the termination of one side), we have to consider whether there is an external interrupt occurring  at time 0 and whether the interrupt assertion turns into the tail assertion at once. Thus, we have the rule:
{\small
\[
\prftree[r]{IntInit}
{
\begin{array}{cc}
     &  \mathsf{sync}(chs,\mathsf{interrupt}(I,e,\{\mathsf{d}\Rightarrow P\},cm),\mathsf{init})(s_1\uplus s_2)\Longrightarrow_a\\
     & \mathsf{interrupt}(I\uplus \mathsf{id\_inv},0,\{\mathsf{d}\Rightarrow \uparrow(e\le0)\bar{\land}\mathsf{sync}(chs,P,\mathsf{init})\}\\
     & \mathsf{rel\_init1}(cm|_{chs^{c}},\mathsf{init}))(s_1\uplus s_2)
\end{array}
}
\]}
\par\noindent
The list function $\mathsf{rel\_init1}$ is obtained from $\mathsf{rel1}$ by replacing $\mathsf{delay}(d,R)$ by $\mathsf{init}$

Synchronization involving recursive assertions is typically very complex, often requiring inductive analysis tailored to specific cases. As such, here we only provide the rule for a specific scenario to facilitate automated implementation.
{\small
\[
\prftree[r]{Rec}{\begin{array}{c}
     \forall\  s_0\,Q.\ \mathsf{sync}(chs,P_1,F_2(Q))(s_0)\Longrightarrow_a \mathsf{false}(s_0)\\
     \forall\  s_0\,Q.\ \mathsf{sync}(chs,F_1(Q),P_2)(s_0)\Longrightarrow_a \mathsf{false}(s_0)\\
     \forall\  s_0.\ \mathsf{sync}(chs,P_1,P_2)(s_0)\Longrightarrow_a P(s_0)\\
     \forall\ s_0\,Q_1\,Q_2.\ \mathsf{sync}(chs,F_1(Q_1),F_2(Q_2))(s_0)\Longrightarrow_a
     F(\mathsf{sync}(chs,Q_1,Q_2))(s_0)
\end{array}
}
{
\begin{array}{c}
     \mathsf{sync}(chs,\mathsf{Rec}\ R_1.\ P_1\bar{\lor} F_1(R_1),\mathsf{Rec}\ R_2.\ P_2\bar{\lor} F_2(R_2))(s_0)  \\
     \Longrightarrow_a \mathsf{Rec}\ R.\ P\bar{\lor} F(R)(s0)
\end{array}
}
\]}
\par\noindent
The first two conditions state that if one side loops while the other doesn't, synchronization results in false. The third condition specifies that when both sides don't loop, synchronization is achieved. The last condition states that if both sides loop, synchronization depends on their outermost loops finishing together. Meeting all four conditions results in a new recursive assertion. This requires consistent recursion counts and simultaneous start and end of each iteration for both sides.

\section{Proof Procedure of Example\ref{example_prop}}
\label{App:prop}

In this section we give the details of the proof procedure of the example in Sect.~\ref{Propertyv} 

According to the rule for recursion assertion, there are three premises to be checked:
{\small
\[
\forall s.\,p(s)\longrightarrow loop(s) \tag{1}
\]
\[
\forall s_0\,s\,tr.\,loop(s_0)\longrightarrow (s_0,s,tr)\models\mathsf{init}\longrightarrow q_1(s)\land \mathsf{trI}(tr,q_2) \tag{2}
\]
\[
\forall Q\, s_0\,s\,tr.\,(\forall s_0\,s\,tr.\,loop(s_0)\longrightarrow (s_0,s,tr)\models Q\longrightarrow q_1(s)\land \mathsf{trI}(tr,q_2))\longrightarrow
\]
\[
loop(s_0)\longrightarrow (s_0,s,tr)\models\mathsf{wait}(\mathsf{id},1,\{\mathsf{d}\Rightarrow Q[x:=x+1]\})\longrightarrow q_1(s)\land \mathsf{trI}(tr,q_2) \tag{3}
\]
}
\par\noindent
(1) is obvious and (2)
is proved by the rule for $\mathsf{init}$ and the trivial fact $\forall s.\,loop(s)\longrightarrow q_1(s)$.
To prove (3), we view (3a) as the assumption and we need to deduce (3b)
{\small
\[
\forall s_0\,s\,tr.\,loop(s_0)\longrightarrow (s_0,s,tr)\models Q\longrightarrow q_1(s)\land \mathsf{trI}(tr,q_2)\tag{3a}
\]
\[
loop(s_0)\longrightarrow (s_0,s,tr)\models\mathsf{wait}(\mathsf{id},1,\{\mathsf{d}\Rightarrow Q[x:=x+1]\})\longrightarrow q_1(s)\land \mathsf{trI}(tr,q_2) \tag{3b}
\]
}
\par\noindent
By applying the rule for $\mathsf{wait}$ to (3b), we obtain the following two conditions. The first one:
{\small
\[
loop(s_0)\land 1>0 \land t\ge 0\land t\le 1 \longrightarrow s=s_0 \longrightarrow q_2(s)
\]
} 
\par\noindent
is also obvious. The second one is:
{\small
\[
loop(s_0)\land 1>0\longrightarrow (s_0,s,tr)\models Q[x:=x+1]\longrightarrow q_1(s)\land \mathsf{trI}(tr,q_2)
\]
}
\par\noindent
By the rule for substitution, we have to prove for all $s_0,s,tr$:
{\small
\[
(\exists v.\, loop[v/x]\land 1>0\land x=v+1)(s_0)\longrightarrow (s_0,s,tr)\models Q\longrightarrow q_1(s)\land \mathsf{trI}(tr,q_2)
\]
}
\par\noindent
To make use of the assumption (3a), we need to prove:
{\small
\[
\forall s.\  (\exists v.\, loop[v/x]\land 1>0\land x=v+1)(s)\longrightarrow loop(s)
\]
}
\par\noindent
which is similar to that the loop invariant is still satisfied after executing one-round loop. Also this logic formula is obviously sound. So far, we have proved the property of this process.

\end{document}